\documentclass[12pt]{article}
\usepackage{graphicx}
\usepackage{times}

\usepackage{epsf}
\def\beq{\begin{equation}}
\def\eeq{\end{equation}}
\def\bea{\begin{eqnarray}}
\def\eea{\end{eqnarray}}

\def\vel{\left|}
\def\ver{\right|}
\def\nnb{\nonumber}

\def\rar{\rightarrow}
\def\nnb{\nonumber}

\def\ba{\begin{array}}
\def\ea{\end{array}}
\def\bea{\begin{eqnarray}}
\def\eea{\end{eqnarray}}

\def\vel{\left|}
\def\ver{\right|}
\def\nnb{\nonumber}

\def\rar{\rightarrow}
\def\nnb{\nonumber}

\def\lla{\left<}
\def\rra{\right>}

\def\es{\!\!\! &=& \!\!\!}

\def\ar{&+& \!\!\!}

\begin{document}

\title{    Double-Lepton Polarization
Asymmetries and Branching Ratio  in $B \rar K_{0}^{*}(1430) l^+ l^- $ transition from
Universal Extra Dimension Model }

\author{ {\small B. B.  \c{S}irvanl{\i}}$^\dag$, {\small K.  Azizi}$^\ddag$, {\small Y.  Ipeko\u glu}$^*$\\
{\small ${}\dag$ Department of Physics, Faculty of Arts and Science, Gazi University, } \\
{\small  Teknikokullar,  06100 Ankara, Turkey}
\\{\small ${}^\ddag$ Physics Division,  Faculty of Arts and Sciences,
Do\u gu\c s University,
 Ac{\i}badem-Kad{\i}k\"oy,}\\{\small34722 Istanbul, Turkey}\\
{\small ${}^*$ Physics Department, Middle East Technical University,
06531 Ankara, Turkey
}} 
\thispagestyle{empty}
\maketitle

\begin{abstract}
We investigate  the $B \rar K_{0}^{*}(1430) l^+ l^- $ transition
in the Applequist-Cheng-Dobrescu  model in the presence of a
universal extra dimension. In particular, we calculate double lepton
polarization asymmetries and branching ratio related to this channel and compare the obtained results with the predictions of the standard model. Our analysis of the considered observables in terms of 
 radius $R$ of the compactified extra-dimension  as the new parameter of the model
show a considerable discrepancy between the predictions
of two models in low $\frac{1}{R}$ values.
\end{abstract}
~~~PACS number(s):12.60.--i, 13.20.--v, 13.20.He

%
\clearpage
\section{Introduction \label{s1}}
The $B \rar K_{0}^{*}(1430) l^+ l^- $ transition proceeds via flavor changing neutral  current (FCNC)  transition of $b\rar s l^+ l^-$ at loop level. Such transition can 
be used in constraining the standard model (SM) parameters as well as gaining useful information about new physics effects such as extra dimensions,
 fourth generation of the quarks,
supersymmetric particles  and  light dark matter , etc.  The SM of particle physics can explain almost all known collider data and is in perfect agreement 
with the experiments so far. However, there are some
problems such as, the origin of the matter in the universe, gauge and fermion mass hierarchy, number of generations, matter-
antimatter asymmetry, unification, quantum gravity and so on, which can not explained
by the SM. Hence, the SM can be thought to be a low energy manifestation of
some underlying more fundamental theory or, to solve the aforementioned problems, some alternative theories are needed.

The extra dimension (ED) model with a flat metric \cite{arkani,Antoniadis,Antoniadis1} or with small compactification radius is one of the alternative theories. The ED is 
categorized as universal extra dimension (UED), where the SM fields containing gauge bosons and fermions can propagate in the extra dimensions
and non-universal extra dimension (NUED), where the gauge bosons propagate into the extra dimensions, but
the fermions are confined to  the usual three spatial dimensions ($D_3$ brane). The  simplest example of the UED where just a single universal extra
dimension is taken into account is called the Appelquist, Cheng and Dobrescu (ACD) model \cite{ACD}. Compared to the SM, this model has one extra parameter called
compactification radius, $R$. Hence, this model is a minimal extension of the SM in $4 + 1$ dimensions with the extra dimension
compactified to the orbifold $S^1 /Z_2$ and the fifth coordinate, $y$ running between $0$ and  $2\pi R$, and $y = 0$ and $y = \pi R$
are fixed points of the orbifold. The zero  modes of fields propagating in the extra dimension correspond the SM particles. The higher modes  with momentum propagating in the extra dimension 
are called Kaluza-Klein (KK)
modes. The mass of KK particles and  interactions among them and also their interactions
with the SM particles are explained in terms of the compactification scale, $1/R$. One of the important properties of the ACD model is conservation of the KK parity, $(-1)^{KK~ number
}$ (for details about the method see also \cite{buras1,buras2,buras3,colangelo0604029}). Such conservation entails the absence of tree level contributions of KK mods to processes occur at
low energies, $\mu\ll \frac{1}{R}$,  requiring the production of a single KK particle from the interaction of th SM  particles.
This allows us to use accurate electroweak measurements
to supply a lower bound to the compactification scale, $\frac{1}{R} \geq (250 -300) ~GeV$ \cite{colangelo0604029,Appelquist}. As these excitations
can affect the loop level processes, especially FCNC transitions, investigation of $B \rar K_{0}^{*}(1430) l^+ l^- $ channel in the framework of the ACD model can be useful for constraining the parameters related
to this new physics scenario. 

The ACD model has been applied widely to calculate many observables related to the radiative and semileptonic decays of hadrons (for some of them 
see for example \cite{buras1,buras2,buras3,colangelo0604029,azizi1,colangelo,aliev,aslambey}. In the present work, we calculate double lepton
polarization asymmetries and branching ratio related to the rare semileptonic $B \rar K_{0}^{*}(1430) l^+ l^- $ transition in terms of 
 radius $R$ of the compactified extra-dimension  as the new parameter of the model  in the framework of the ACD model. We compare
 the obtained results with the predictions of the standard model. The outline of the paper is as follows. In  section 2, we introduce the effective Hamiltonian responsible for the $b\rar s l^+ l^-$ transition. Using the effective Hamiltonian, we obtain the branching ratio as well as the various related double lepton
polarization asymmetries in terms of form factors also in this section. Using the fit parametrization of the form factors obtained using  QCD sum rules, we numerically analyze the considered observables in section 3.
 This section also includes a comparison of the results obtained in ACD model with that of predicted by the SM and our discussions.

\section{Branching ratio and double lepton
polarization asymmetries in  $B \rar K_0^{*} l^+ l^- $  transition  \label{s2}}
At quark level, the $B \rar K_0^{*} l^+ l^- $  transition proceed via FCNC transition of the $b \rar s l^+ l^- $. The  effective Hamiltonian 
responsible for this transition at quark level can be written as:

 \bea \label{e8401} {\cal H}^{eff} &=& {G_F \alpha_{em} V_{tb}
V_{ts}^\ast \over 2\sqrt{2} \pi} \Bigg[ C_9^{eff} 
\bar{s}\gamma_\mu (1-\gamma_5) b \, \bar{\ell} \gamma^\mu \ell +
C_{10}  \bar{s} \gamma_\mu (1-\gamma_5) b \, \bar{\ell}
\gamma^\mu
\gamma_5 \ell \nnb \\
&-&  2 m_b C_7^{eff}  {1\over q^2} \bar{s} i \sigma_{\mu\nu}
(1+\gamma_5) b \, \bar{\ell} \gamma^\mu \ell \Bigg]~, \eea

where $G_F$ is the Fermi constant, $\alpha_{em}$  is the fine structure
constant at Z mass scale, $V_{ij}$ are elements of the
Cabibbo-Kobayashi-Maskawa (CKM) matrix and 
$C_7^{eff}$, $C_9^{eff}$ and $C_{10}$ are the Wilson coefficients, which are the main source of the deviation of  the ACD and  SM models predictions on the considered observables. 
The Wilson coefficients can be expressed in terms of the periodic
functions, $F(x_{t},1/R)$ with $x_{t}=m_{t}^{2}/M_{W}^{2}$ and $m_t$ being 
the top quark mass. Similar to the mass of the KK particles described in terms of the zero modes $(n = 0)$
correspond to the ordinary particles of the SM and extra parts coming from the ACD model, the functions, $F(x_{t},1/R)$ are also  written in terms of the corresponding SM functions, $F_0 (x_t )$ and extra parts which are functions of the 
compactification factor, $1/R$, i.e., 
\bea F(x_t,1/R)=F_0(x_t)+\sum_{n=1}^{\infty}F_n(x_t,x_n),
\label{functions} \eea where $x_n=\displaystyle{m_n^2 \over
M_W^2}$ and $m_n=\displaystyle{n \over R}$. The Glashow-Illiopoulos-Maiani (GIM) mechanism guarantees the finiteness of the functions, $F(x_t,1/R)$ and satisfies the 
condition, $F(x_t,1/R)\rar F_0(x_t)$, when $R\rar 0$. As far as $1/R$ is taken in the order of a few hundreds of $GeV$, these functions and as a result, the Wilson coefficients  
differ considerably from the SM values. For explicit expressions of the Wilson coefficients in ACD model see \cite{buras1,buras2,colangelo0604029}.

 To obtain the amplitude for the 
 $B \rar
K_0^{*} l^+ l^- $ transition, we need to sandwich the effective Hamiltonian between the initial and final states. As a result of this procedure,  the matrix elements, $\lla
K_0^\ast \vel \bar{s}\gamma_\mu (1-\gamma_5)b \ver B \rra$ and
$\lla K_0^\ast \vel \bar{s} i \sigma_{\mu\nu} q^\mu (1+\gamma_5)b
\ver B \rra$ are obtained which should be calculated in terms of some form factors. Due to the parity considerations, the vector ($\overline{s}\gamma_{\mu}b$) and tensor ($\overline{s}i \sigma_{\mu\nu}
q^{\nu}b$) parts of the transition current have no contributions. The matrix elements related to the axial-vector and pseudo-tensor parts of the transition currents are parameterized 
in terms of the form factors, $f_{+}$, $f_{-}$, and  $f_{T}$ in the following
way:

\bea \lla
K_0^\ast (p')\vel \bar{s}\gamma_\mu \gamma_5b \ver B(p) \rra=f_{+}(q^2)\mathcal{P}_{\mu}+f_{-}(q^2)q_{\mu}\eea

\bea\lla K_0^\ast(p') \vel \bar{s} i \sigma_{\mu\nu} q^\mu \gamma_5b
\ver B(p) \rra = \frac{f_{T}(q^2)}{m_B +
m_{K_{0}^{*}}}[\mathcal{P}_{\mu}q^2-(m_B^2 -
m^{2}_{K_0^*})q_{\mu}]\eea where ${\cal P} =
p+p'$ and $q =
p-p'$. These form factors have been calculated in \cite{azizi0710.1508} in the framework of the three-point QCD sum rules. The fit parametrization of the form factors is given as:

\bea \label{e8425} f_i(\hat{s}) = {f_i(0)\over 1 - a_i \hat{s} +
b_i \hat{s}^2}~, \eea where $i=+$, $-$ or $T$ and $\hat{s} =
q^2/m_B^2$. The values of the parameters $f_i(0)$, $a_i$ and $b_i$
are given in Table 1.
\begin{table}[h]
\renewcommand{\arraystretch}{1.5}
\addtolength{\arraycolsep}{3pt}
$$
\begin{array}{|l|ccc|}
\hline & f_i(0) & a_i & b_i \\ \hline f_+ &
\phantom{-}0.31 \pm 0.08 & 0.81 & -0.21 \\
f_- &
-0.31\pm 0.07 & 0.80 & -0.36 \\
f_T & -0.26\pm 0.07 & 0.41 & - 0.32 \\ \hline
\end{array}
$$
\caption{parameters entering the fit parametrization of the form factors for $B \rar K_0^\ast \ell^+ \ell^-$ transition.}
\renewcommand{\arraystretch}{1}
\addtolength{\arraycolsep}{-3pt}
\end{table}

Now, we proceed to calculate the differential decay rate for the considered transition. Using the amplitude and definition of the transition matrix elements in terms of the  form factors, we get 
the following expression for the
$1/R$-dependent differential decay rate:
\bea
\label{e8422}
{d \Gamma \over d\hat{s}}(\hat{s},1/R) \es
{G^2 \alpha_{em}^2 m_B^5 \over 3072 \pi^5} \vel V_{tb} V_{ts}^\ast \ver^2
v \sqrt{\lambda(1,\hat{m}_{K_0^\ast}^2,\hat{s})} \Bigg\{ \Bigg[
\vel C_9^{eff}(\hat{s},1/R) f_+(\hat{s}) + {2 \hat{m}_b \over 1 + 
\hat{m}_{K_0^\ast}} C_7^{eff}(1/R) f_T(\hat{s}) \ver^2 \nnb \\
\ar \vel C_{10} (1/R)f_+(\hat{s}) \ver^2 \Bigg] (3-v^2)
\lambda(1,\hat{m}_{K_0^\ast}^2,\hat{s}) +
12 \hat{m}_\ell^2 \Big[ (2 + 2 \hat{m}_{K_0^\ast}^2 -\hat{s}) 
\vel f_+(\hat{s}) \ver^2 \nnb \\ 
\ar 2 (1-\hat{m}_{K_0^\ast}^2) {\rm Re} [f_+(\hat{s}) f_-^\ast(\hat{s})] + 
\hat{s} \vel f_- (\hat{s})\ver^2  \Big]\vel C_{10}(1/R) \ver^2
\Bigg\}~,
\eea
where, 
$
 v = \sqrt{1 - {4 \hat{m}_\ell^2 \over
\hat{s}}},~ \hat{m}_b = {m_b \over m_B},~
\hat{m}_\ell = {m_\ell \over m_B},~
\hat{m}_{K_0^\ast} = {m_{K_0^\ast} \over m_B}$ and $ \lambda(a,b,c) = a^2+b^2+c^2-2ab-2ac-2bc$ is the usual triangle function.
 Integrating out the above equation in the allowed physical region of the $\hat{s}$ ($4\hat{m}_\ell^2\leq \hat{s}\leq(1-\hat{m}_{K_0^\ast})^2$), one can get the $1/R$-dependent total decay rate and branching ratio.

At the end of this section, we focus our attention to obtain the double-lepton polarization asymmetries. We calculate these asymmetries when polarizations of both leptons 
 simultaneously are considered.
Using the definitions of the double-lepton polarization asymmetries expressed in \cite{veli,veli0710.2619,Fukae}, we obtain the $1/R$-dependent polarizations,
\begin{eqnarray}
P_{LL}(\hat{s},1/R)&=&\frac{-4m_B^2}{3\Delta(\hat{s},1/R)}Re[-24m_B^2\hat{m}_{l}^2(1-\hat{r}_{K_{0}^{*}})C^{*}D+\lambda
m_B^2(1+v^2)|A|^2\\&-& 12m_B^2 \hat{m}_{l}^2\hat{s}|D|^2
+m_B^2|C|^2(2\lambda-(1-v^2)(2\lambda+3(1-\hat{r}_{K_{0}^{*}})^2))],\\
P_{LN}(\hat{s},1/R)&=&\frac{-4\pi m_{B}^{3} \sqrt{\lambda
\hat{s}}}{\hat{s}\Delta(\hat{s},1/R)}
Im[-m_{B}\hat{m}_{l}\hat{s}A^{*}D -m_{B}\hat{m}_{l}(1-\hat{r}_{K_{0}^{*}})A^{*}C],\\
P_{NL}(\hat{s},1/R)&=&-P_{LN}(\hat{s},1/R),\\
P_{LT}(\hat{s},1/R)&=&\frac{4\pi
m_{B}^3\sqrt{\lambda\hat{s}}}{\hat{s}\Delta(\hat{s},1/R)}Re[m_{B}\hat{m}_{l}v(1-\hat{r}_{K_{0}^{*}})|C|^2+m_{B}\hat{m}_{l}v\hat{s}C^{*}D],\\
P_{TL}(\hat{s},1/R)&=&P_{LT}(\hat{s},1/R),\\
P_{NT}(\hat{s},1/R)&=&-\frac{8m_B^2v}{3\Delta(\hat{s},1/R)}Im[2\lambda m_{B}^2A^{*}C],\\
P_{TN}(\hat{s},1/R)&=&-P_{NT}(\hat{s},1/R),\\
P_{TT}(\hat{s},1/R)&=&\nonumber\frac{4m_B^2}{3\Delta(\hat{s},1/R)}Re[-24
m_{B}^2\hat{m}_{l}^2(1-\hat{r}_{K_{0}^{*}})C^{*}D-\lambda
m_{B}^2(1+v^2)|A|^2-12m_{B}^2\hat{m}_{l}^2\hat{s}|D|^2\\&+&m_{B}^2|C|^2\{2\lambda-(1-v^2)(2\lambda+3(1-\hat{r}_{K_{0}^{*}})^2)\}],\\
P_{NN}(\hat{s},1/R)&=&\nonumber\frac{4m_B^2}{3\Delta(\hat{s},1/R)}Re[24
m_{B}^2\hat{m}_{l}^2(1-\hat{r}_{K_{0}^{*}})C^{*}D-\lambda
m_{B}^2(3-v^2)|A|^2+12m_{B}^2\hat{m}_{l}^2\hat{s}|D|^2\\&+&m_{B}^2|C|^2\{2\lambda-(1-v^2)(2\lambda-3(1-\hat{r}_{K_{0}^{*}})^2)\}]
\end{eqnarray}
where, $L$, $N$ and $T$ stand for the longitudinal,
normal and transversal polarizations, respectively,
 $\hat{r}_{K_{0}^{*}}=\hat{m}_{K_0^\ast}^2 $, $\lambda=\lambda(1,\hat{r}_{K_{0}^{*}},\hat{s})$ and
\bea \Delta(\hat{s},1/R)&=&\nonumber\frac{4m_B^2}{3}Re[24
m_{B}^2\hat{m}_{l}^2(1-\hat{r}_{K_{0}^{*}})D^{*}C+\lambda
m_{B}^2(3-v^2)|A|^2+12m_{B}^2\hat{m}_{l}^2\hat{s}|D|^2\\
&+&
m_{B}^2|C|^2\{2\lambda-(1-v^2)(2\lambda-3(1-\hat{r}_{K_{0}^{*}})^2)\}]\nnb \\
A=A(\hat{s},1/R) &=&2C_{9}^{eff} (\hat{s},1/R) f_+(\hat{s})
-
4C_{7}^{eff}(1/R)(m_b+ms) \frac{f_T(\hat{s})}{m_B+m_{K_{0}^{*}}} , \nnb \\
B=B(\hat{s},1/R) &=& 2C_{9}^{eff} (\hat{s},1/R) f_-(\hat{s}) +4C_{7}^{eff}(1/R)(m_b+ms) \frac{f_T(\hat{s})}{(m_B+m_{K_{0}^{*}})\hat{s}m_B^2}(m_B^2-m_{K_{0}^{*}}^2) , \nnb \\
C=C(\hat{s},1/R) &=& 2C_{10}(1/R) f_+(\hat{s}) ,\nnb \\
D=D(\hat{s},1/R) &=& 2C_{10}(1/R)f_-(\hat{s}) ~. \eea

\section{Numerical results }
In this section, we numerically analyze the expressions of the branching ratio and double-lepton polarization asymmetries and discuss their dependence and sensitivity  on the compactification factor, $1/R$.
Some input parameters  of the SM used in the numerical analysis  are: $m_{t}=167~GeV$,
$m_{W}=80.4~GeV$, $m_{Z}=91.18~GeV$, $m_{c}=1.46~GeV$, $m_{b}=4.8~GeV$, $m_{u}=0.005~GeV$, $m_{B}=5.28~GeV$,
$m_{K_{0}^{*}}=1.425~GeV$, $sin^2 \theta_W =0.23$, $\alpha_{em}=\frac{1}{137}$, $\alpha_s(m_Z)=0.118$,  $| V_{tb} V_{ts}^* |= 0.041$, $G_F = 1.167\times10^{-5}~GeV^{-2}$,
$m_e = 5.1\times10^{-4}~GeV$, $m_\mu=0.109~GeV$, $m_\tau=1.784~GeV$, and $\tau_B=1.525\times10^{-12}s$. As we previously mentioned, the branching ratio is obtained integrating the differential decay rate over
$\hat{s}$ in the physical region of the square of the momentum transfered, $q^2$, hence the obtained expression for the branching ration only depends on the compactification factor.  In Fig. 1,
we present the dependence  branching ratio of the
$B \rar K_{0}^{*} l^+ l^- $ transition on compactification parameter, $1/R$ in the interval, $200~GeV\leq1/R\leq1000~GeV$ for different leptons.
%
\begin{figure}
\includegraphics[scale=0.5]{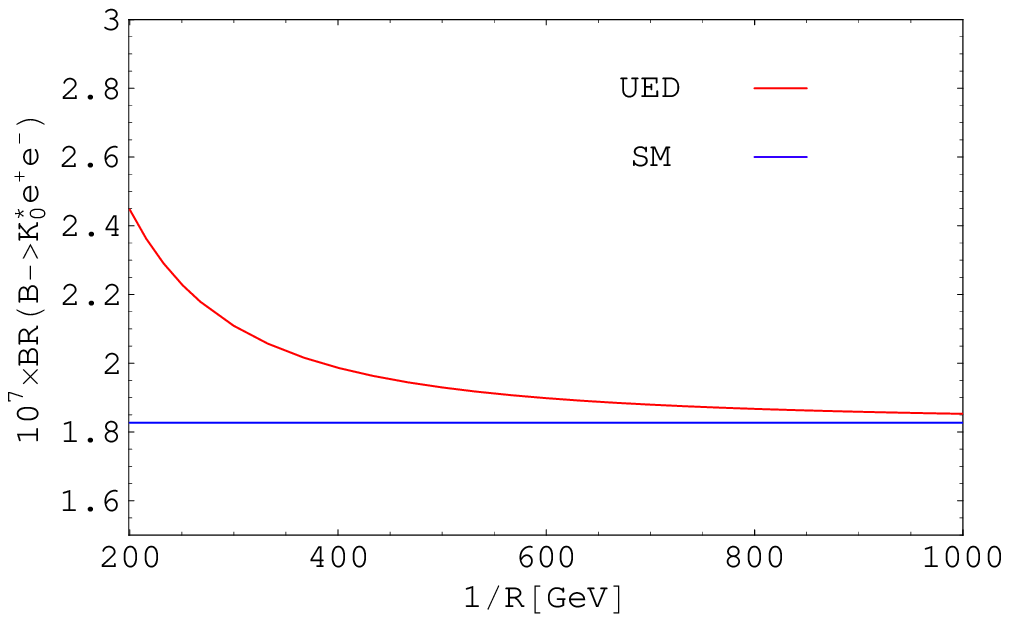}
\includegraphics[scale=0.5]{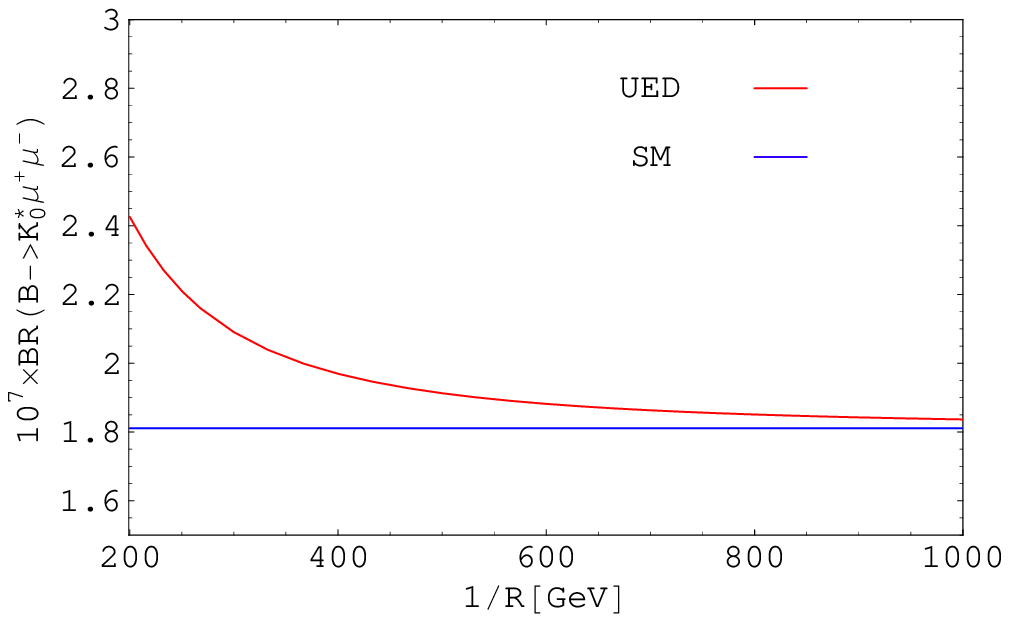}
\includegraphics[scale=0.5]{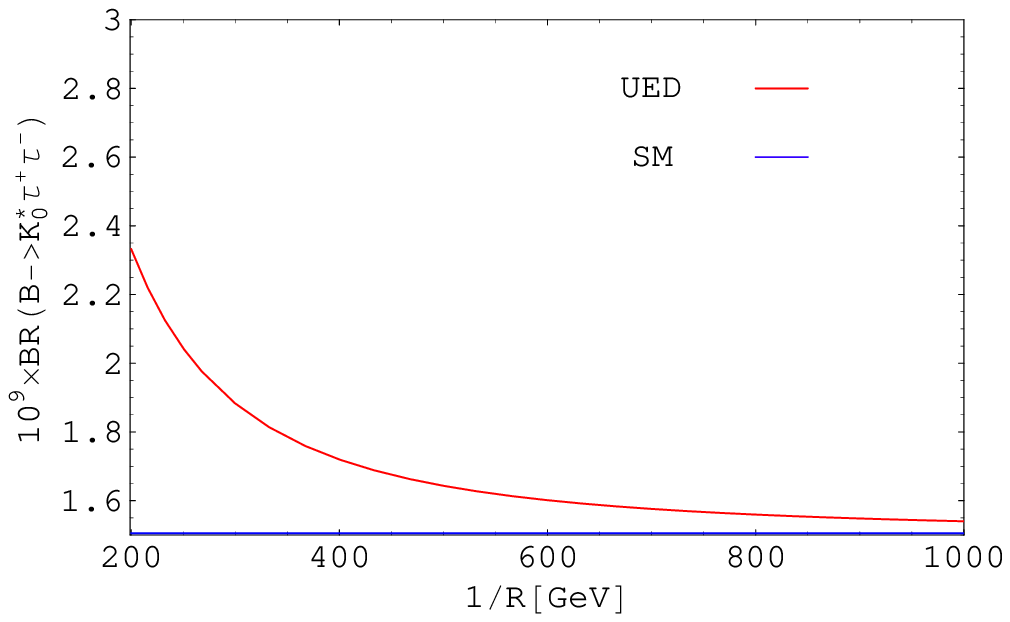}
\caption{ The dependence of the branching ratio for $B \rar
K_{0}^{*} l^+ l^- $ on the compactification factor, $1/R$ for different leptons.
\label{fig1}}
\end{figure}

 From this figure, we deduce the following results:
\begin{itemize}
 \item There are  considerable discrepancies between the predictions of the ACD and SM models for  low  values of the compactification factor, $1/R$. As  $1/R$ increases, the difference between the 
predictions of the two models tends to  diminish. The result of ACD  approaches  the  result of  SM for higher values of   $1/R$ ($1/R\simeq1000~GeV$). Such a discrepancy at low values of $1/R$ can be 
 a signal for the  existence of extra dimensions.
\item As it is expected, an increase in the lepton mass  results in a decrease in the branching ratio. The  branching ratios for the $e$ and $\mu$ are approximately the same.
\item The order of magnitude of the branching ratio, specially for the $e$ and $\mu$, depicts a possibility to study such channels at the LHC.
\end{itemize}
\begin{figure}
\includegraphics[scale=0.5]{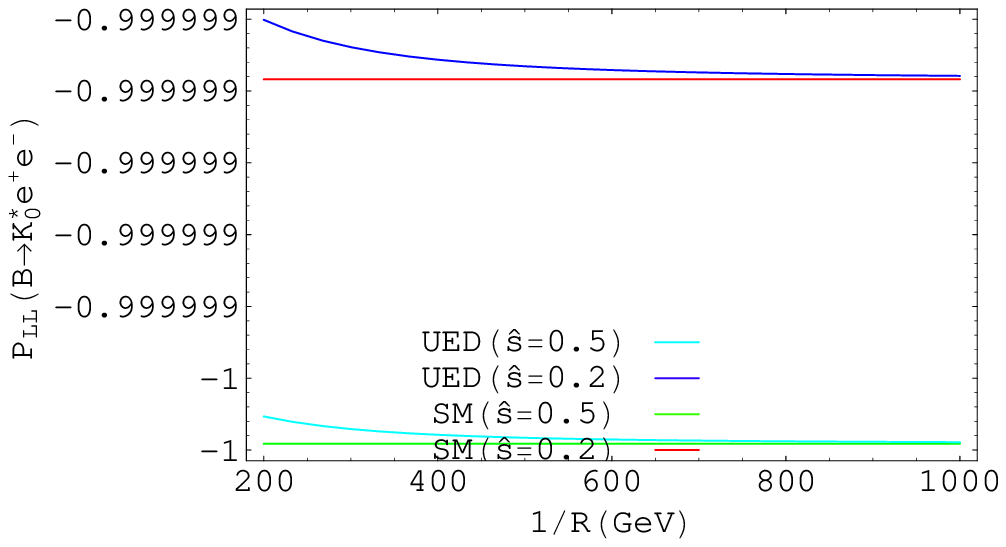}
\includegraphics[scale=0.5]{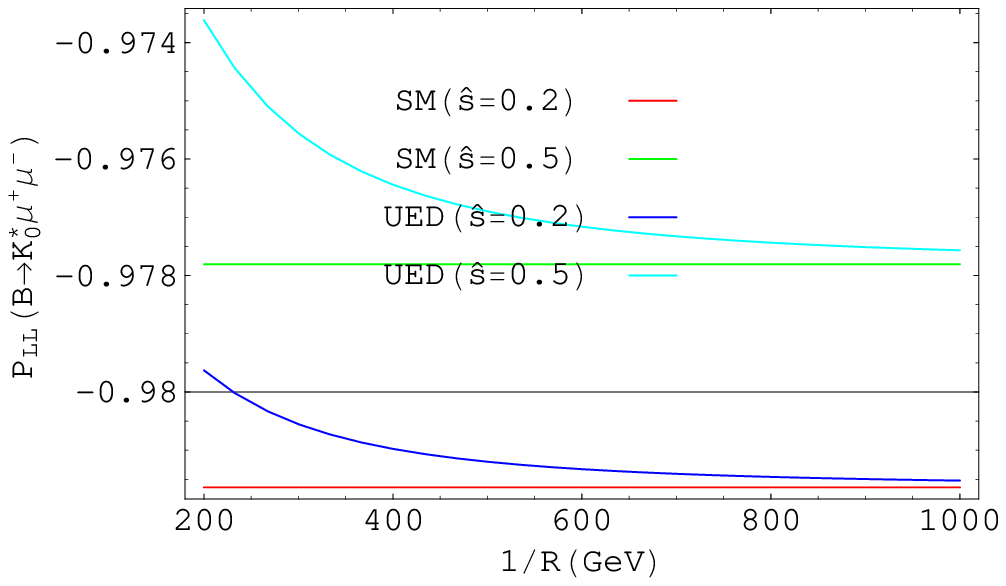}
\includegraphics[scale=0.5]{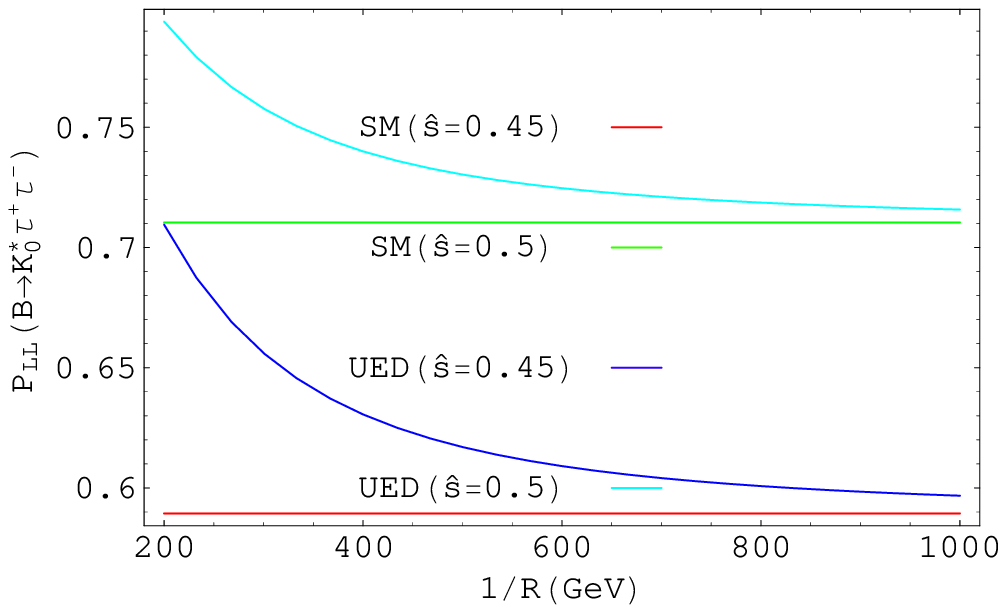}
\caption{ The dependence of the $P_{LL}$ polarization in two models for $B
\rar K_{0}^{*} l^+ l^- $ on the compactification factor, $1/R$ at different fixed values of  $\hat{s}$ and different leptons.
\label{fig3}}
\end{figure}
\begin{figure}
\includegraphics[scale=0.5]{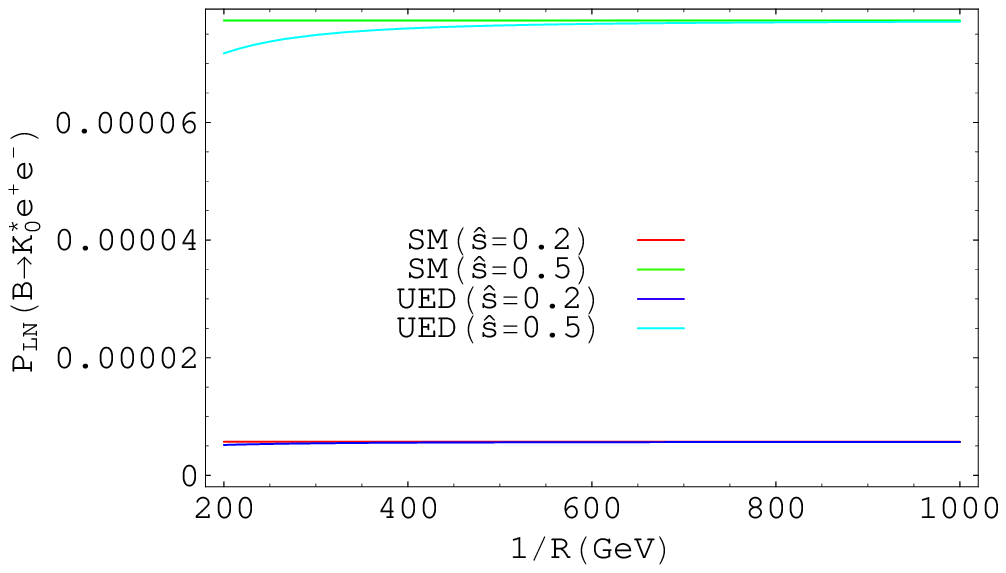}
\includegraphics[scale=0.5]{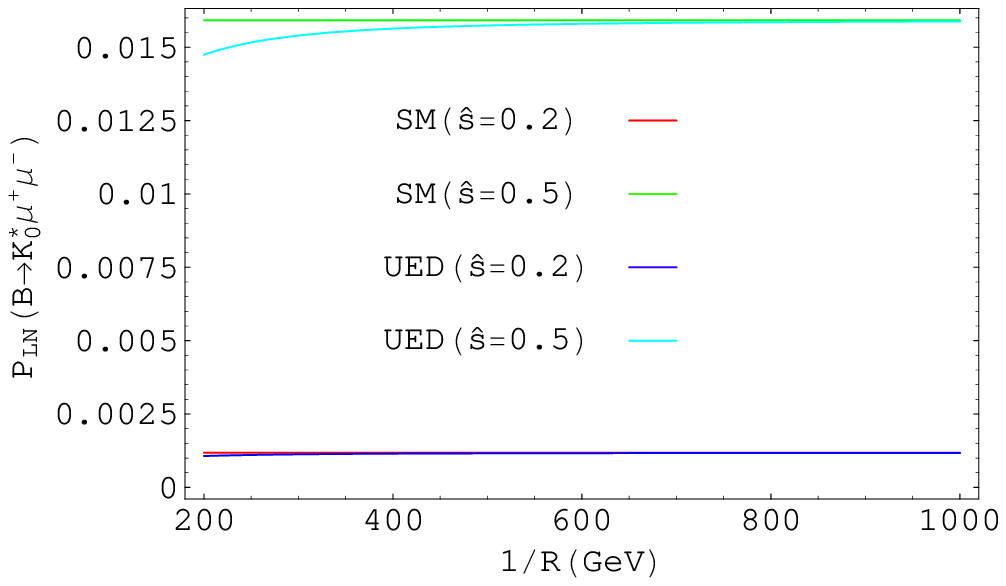}
\includegraphics[scale=0.5]{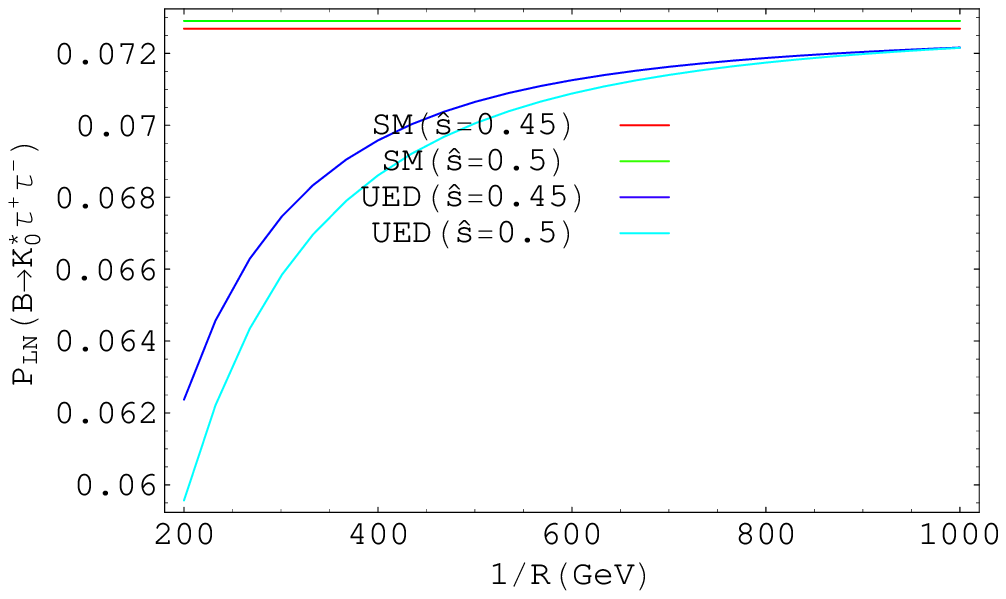}
\caption{ The same as Fig. \ref{fig3}, but for  $P_{LN}$ polarization.
\label{fig5}}
\end{figure}
\begin{figure}
\includegraphics[scale=0.5]{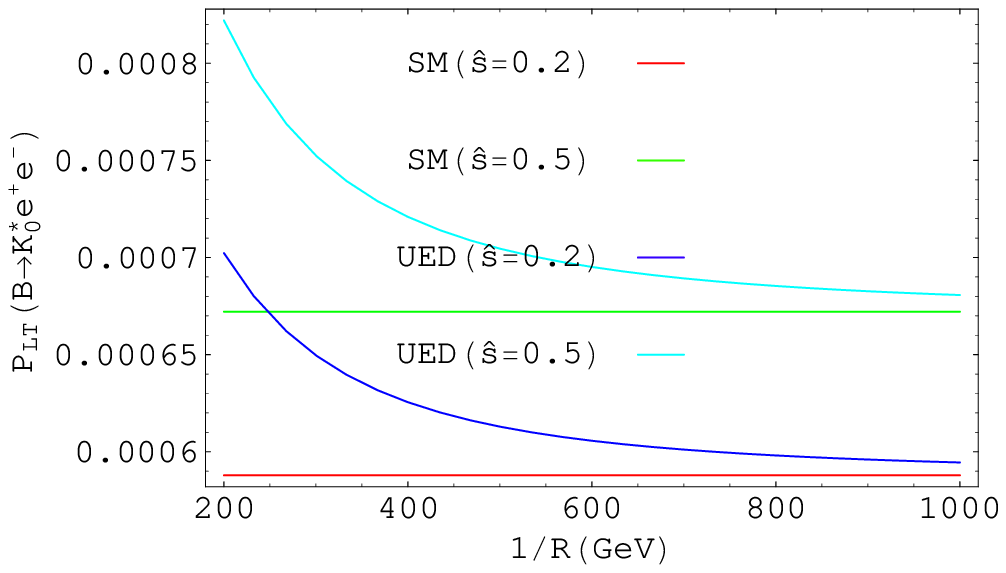}
\includegraphics[scale=0.5]{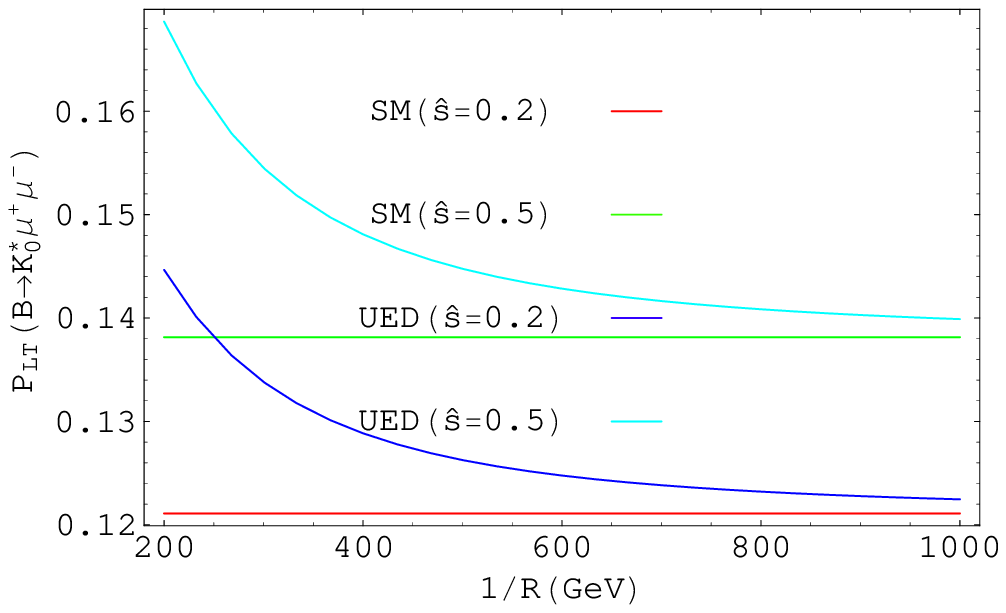}
\includegraphics[scale=0.5]{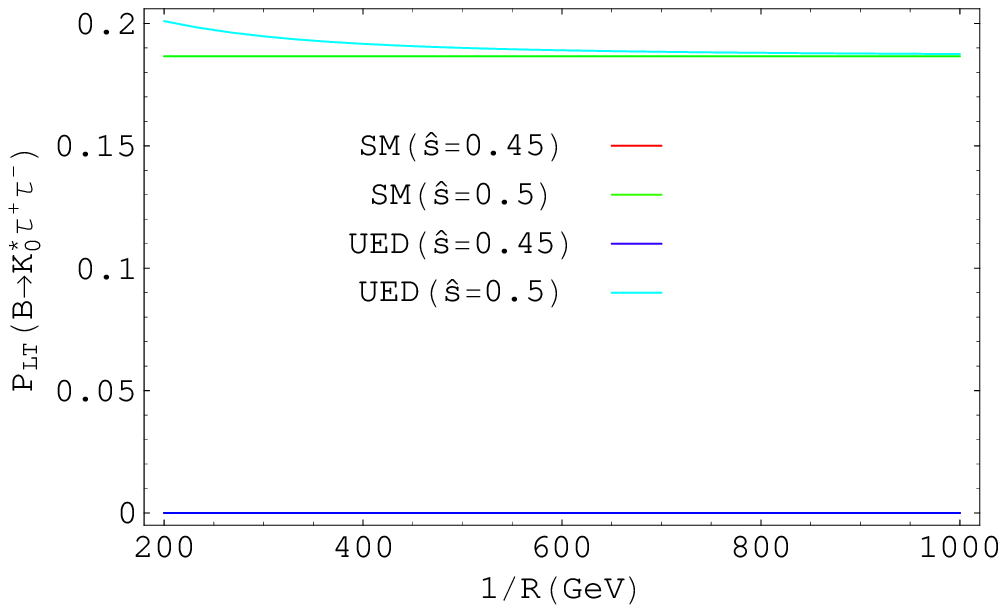}
\caption{  The same as Fig. \ref{fig3}, but for  $P_{LT}$ polarization.
\label{fig7}}
\end{figure}
\begin{figure}
\includegraphics[scale=0.5]{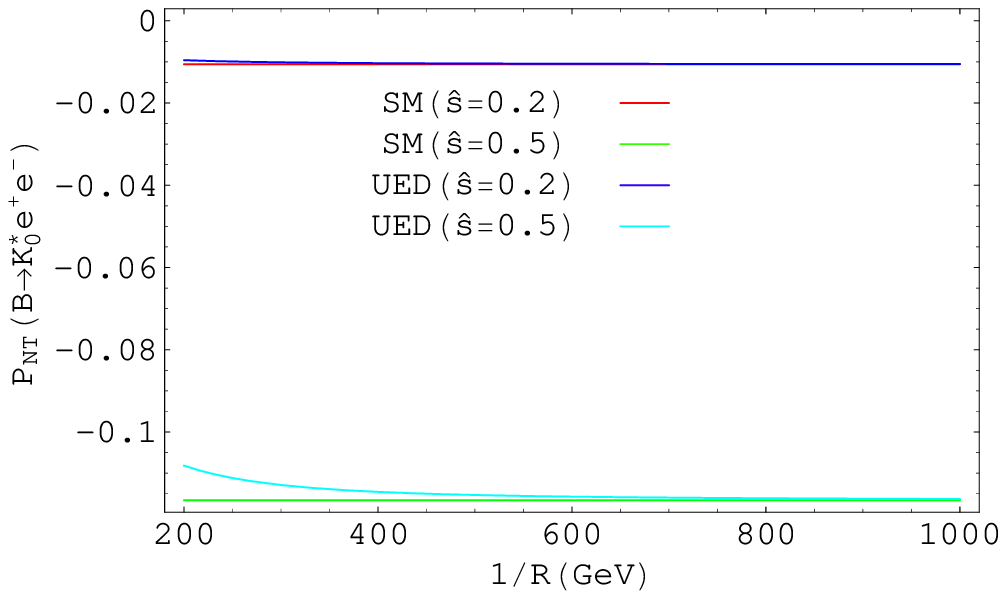}
\includegraphics[scale=0.5]{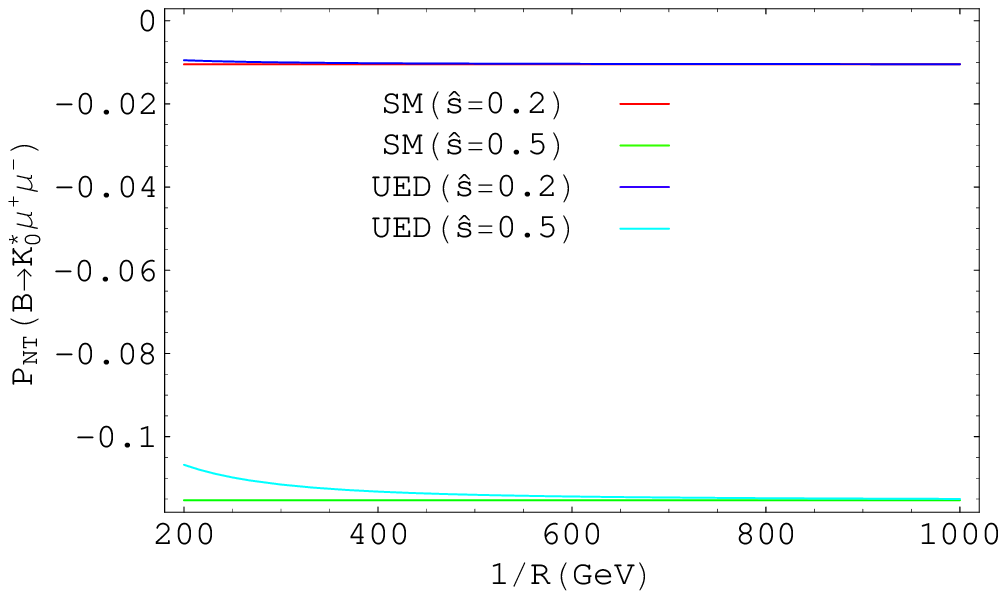}
\includegraphics[scale=0.5]{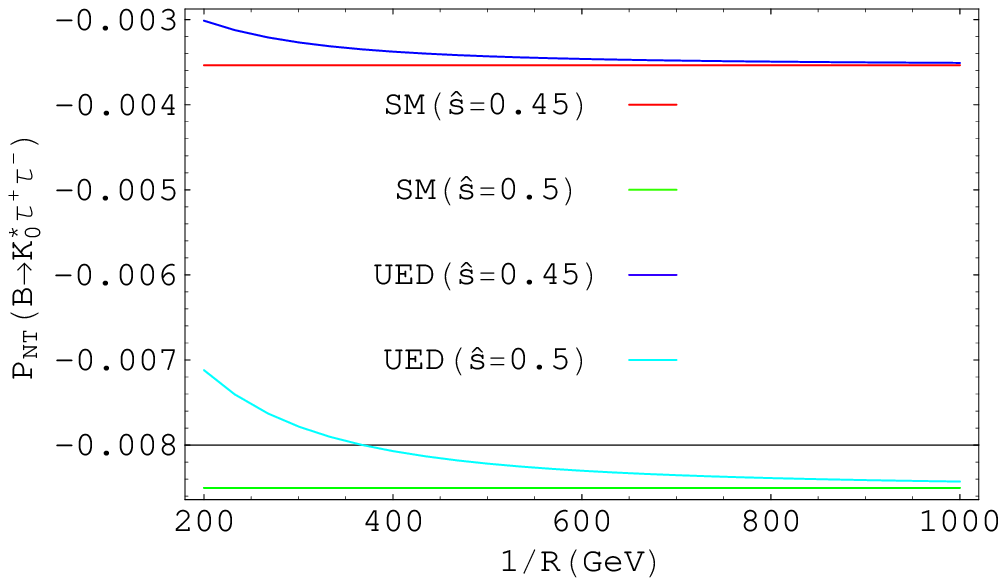}
\caption{ The same as Fig. \ref{fig3}, but for  $P_{NT}$ polarization.
\label{fig9}}
\end{figure}
\begin{figure}
\includegraphics[scale=0.5]{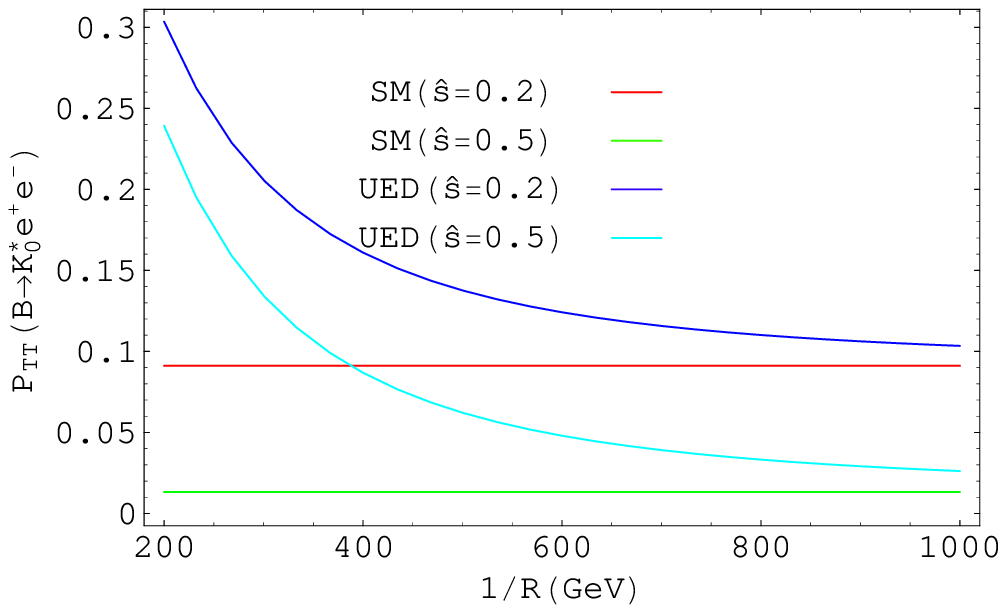}
\includegraphics[scale=0.5]{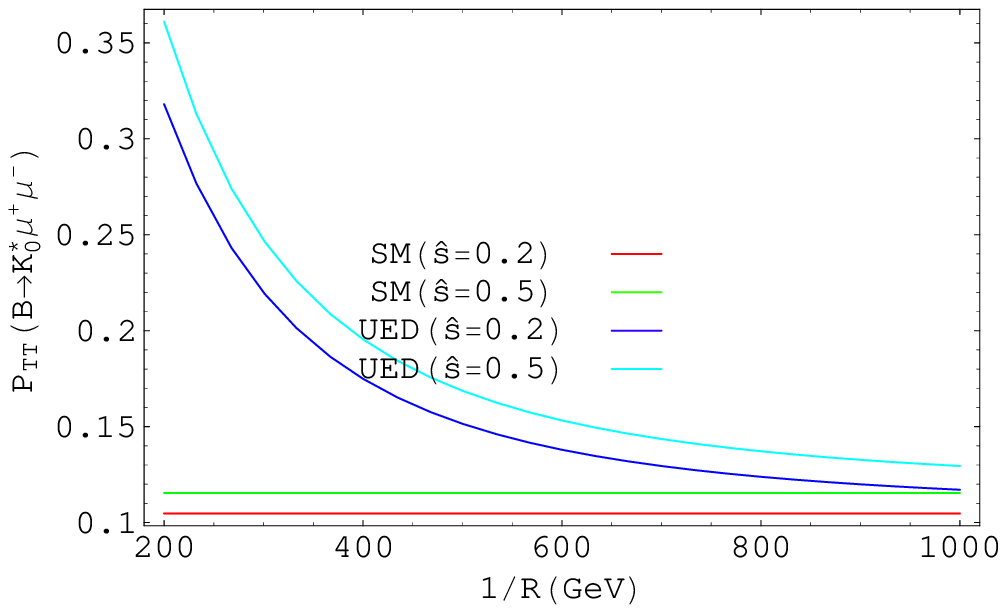}
\includegraphics[scale=0.5]{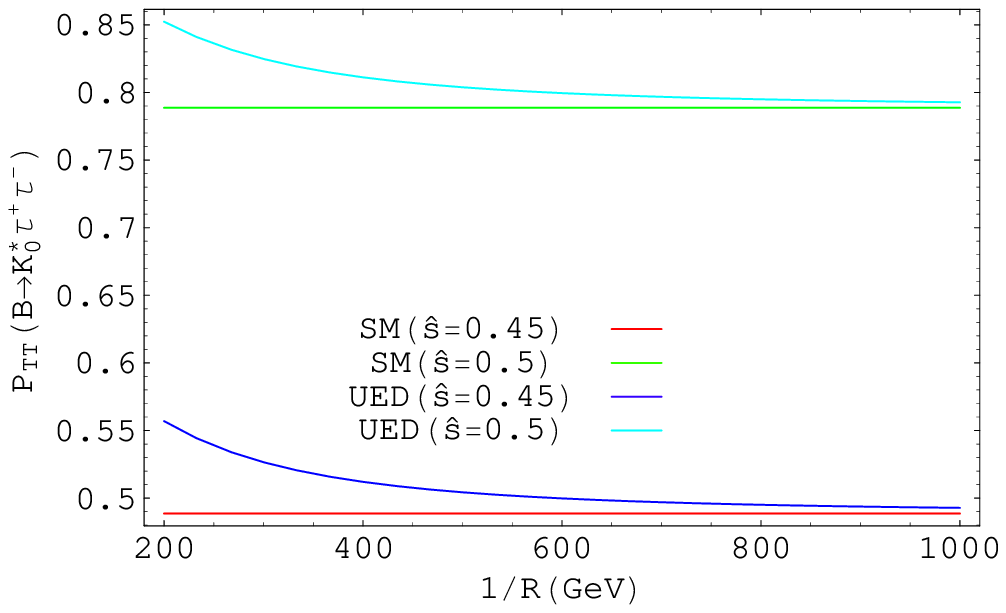}
\caption{ The same as Fig. \ref{fig3}, but for  $P_{TT}$ polarization.
\label{fig11}}
\end{figure}
\begin{figure}
\includegraphics[scale=0.5]{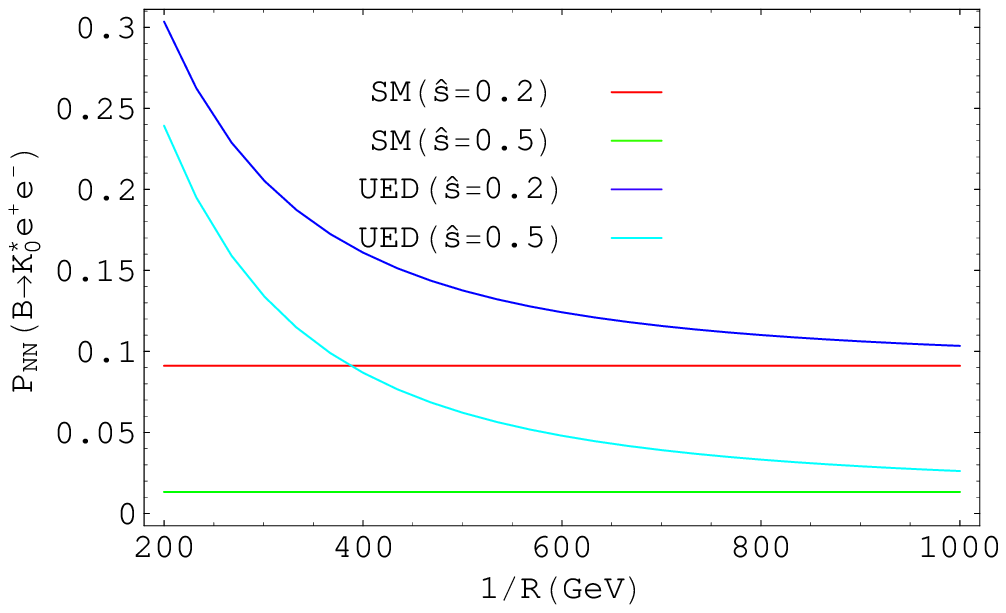}
\includegraphics[scale=0.5]{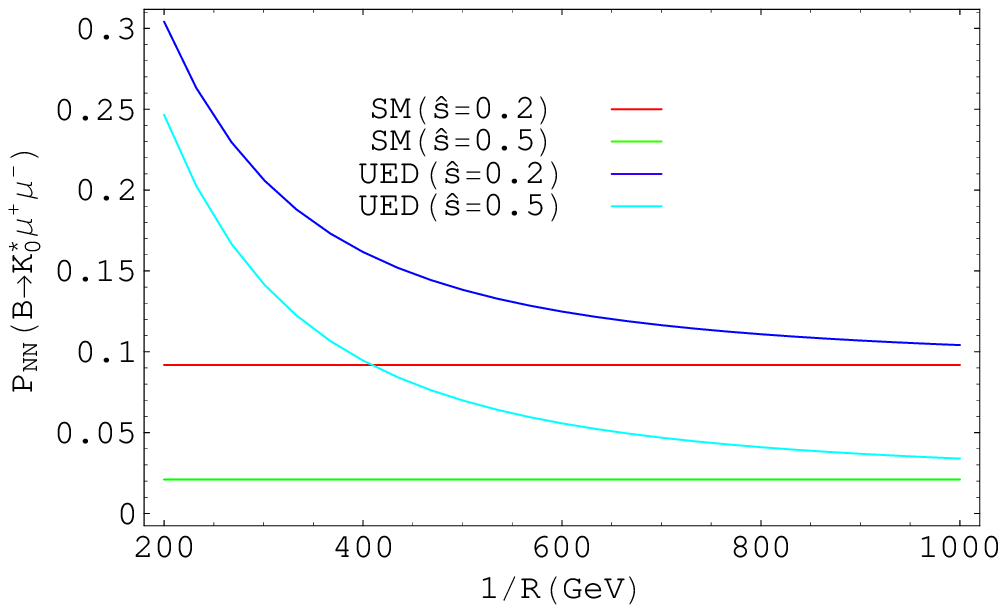}
\includegraphics[scale=0.5]{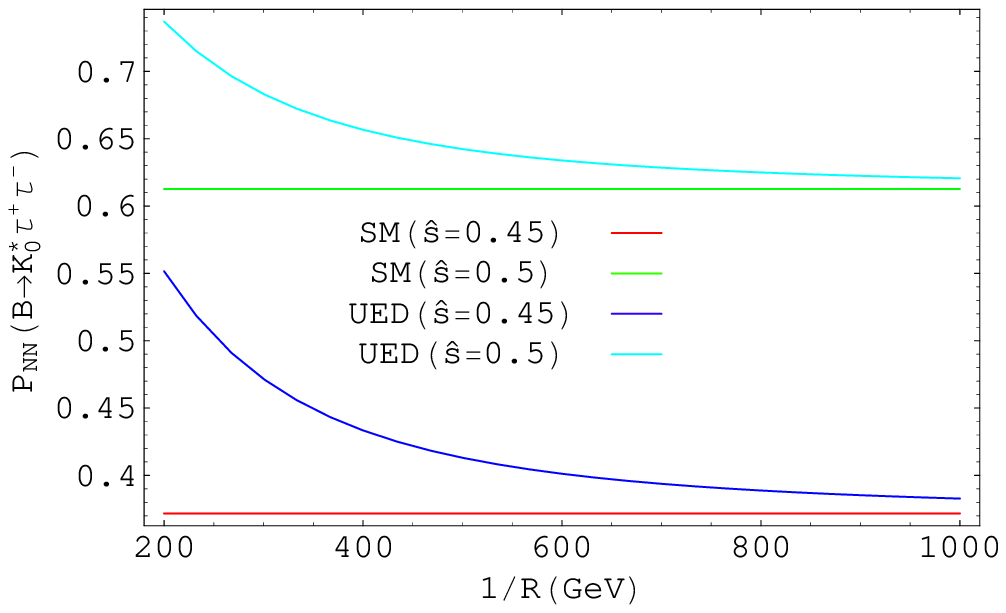}
\caption{The same as Fig. \ref{fig3}, but for  $P_{NN}$ polarization. 
\label{fig13}}
\end{figure}
%
 Now, we proceed to show the results of the double-lepton polarization asymmetries. As it is clear from their explicit expressions, they  depend on both the compactification factor, $1/R$ and the $\hat{s}$.
The dependence of different polarization asymmetries on the compactification factor, $1/R$ at different fixed values of  $\hat{s}$ and different leptons are shown in Figs. \ref{fig3}-\ref{fig13}. A quick 
glance at these figures leads to the following conclusions:
\begin{itemize}
\item As it is expected, all polarization asymmetries lie between $-1$ and $1$. There are also  discrepancies between the predictions of two models for small $1/R$ 
values, except the $P_{LL}$, $P_{LN}$ and $P_{NT}$ for $e$ and $\mu$ cases for
 which the differences between the ACD and SM models are very small.
 At high values of  $1/R$, two models have approximately the same predictions.
\item All double-lepton polarization asymmetries  have the same sign for all leptons, except the $P_{LL}$ at which the sign for the $\tau$ mode is 
different than those for the $e$ and $\mu$ cases. 
\item In contrast with the branching ratios, some of the  polarization asymmetry predictions are different for the $e$ and $\mu$ cases.
\item In $P_{NN}$ for $e$ and $\mu$ and  $P_{TT}$  for $e$, the SM gives zero for $\hat{s}=0.5$, while we see  considerable nonzero values for them in ACD model at low $1/R$ values.
\end{itemize}
We depict the dependence of the different double-lepton polarization asymmetries on $\hat{s}$ at different fixed values of  $1/R$ and the SM for different leptons
 in Figs. \ref{figbelma}-\ref{fig12}. From these figures, we infer the following information:
\begin{itemize}
 \item The longitudinal-longitudinal polarization asymmetry, $P_{LL}$, remains approximately unchanged in whole physical region  except the end points for the $e$ and $\mu$. This asymmetry grows 
as
$\hat{s}$ increases and reaches  its maximum at the upper bound of the allowed physical region for $\tau$.
\item The $P_{LN}$ is zero in the interval, $4 \hat{m}_l^2\leq \hat{s}\leq0.37$, but after $\hat{s}=0.37$ it starts to increase up to the upper limit for $\hat{s}$ ($(1-\hat{m}_{K_0^\ast})^2$) for $e$ and $\mu$. For $\tau$
case, this asymmetry remains approximately unchanged in the interval $4 \hat{m}_l^2\leq \hat{s}\leq0.50$, but after this point it starts to diminish and becomes zero at the endpoint.
\item The $P_{LT}$ slightly decreases as $\hat{s}$ increases for the $e$ and $\mu$ cases and has a very small value for the $e$ case compared to that of the  $\mu$. This asymmetry for $\tau$,
 first increases  then it
 decreases after reaching a maximum as  $\hat{s}$ increases in the allowed physical region.
\item The normal-transversal polarization asymmetry, $P_{NT}$ remain also unchanged   in the interval, $4 \hat{m}_l^2\leq \hat{s}\leq0.37$, for $e$ and $\mu$, but it grows after this interval 
and has negative sign. In the case
 of $\tau$, this asymmetry also has negative sign and it increases to reach a maximum then decreases as $\hat{s}$ increases.
\item For the $e$ and $\mu$, the $P_{TT}$ and $P_{NN}$ start to decrease as $\hat{s}$ increases. The values of these asymmetries in the  SM and higher values of the compactification
 factor become zero around   $\hat{s}=0.37$ then they start to increase as $\hat{s}$ increases. They have minimums at  low  $1/R$ values  at the same $\hat{s}$.
 For $\tau$ case, the $P_{TT}$ and $P_{NN}$ grow as $\hat{s}$ increases and the $P_{NN}$ reaches  its maximum at the upper bound.

\end{itemize}
At the end of this section, we would like to quantify  the uncertainties of our predictions associated with the errors of the hadronic form factors. In this connection,
 we show the dependence of the differential branching ratios for $\mu$ and $\tau$  on $\hat{s}$ in Fig. \ref{difbr} and dependence of $P_{LL}$,  $P_{TT}$ and $P_{NN}$ on $\hat{s}$ 
only for $\mu$  in Fig. \ref{error}. These figures contain our predictions (a) in ACD model at $1/R=200~GeV$ when the errors presented in Table 1 are added to the central values of the form factors,
 (b) the same model and $1/R$, but when the errors of form factors are subtracted from the central values and (c) in SM when  central  values of the form factors are considered.
From Fig. \ref{difbr}, 
we see that the results of SM lies between the cases (a) and (b) but close to the case (b). In the case of $\mu$, the maximum deviation from the SM is in
 lower values of $\hat{s}$ and  belongs to the case (a) and has the value about two times grater than that of the SM. For $\tau$ case, the maximum deviation of the ACD prediction
lies at  middle of the allowed physical region for the  $\hat{s}$. At this point, the ACD prediction in the case (a) is approximately six times greater than the SM prediction. The 
similar deviations from the SM model has also been observed for instance in \cite{aslam} for $\Lambda_b \to \Lambda l^{+} l^{-}$, but at higher values of $\hat{s}$. Figure,  \ref{error}
depicts an interesting observation. The ACD model in the cases (a) and (b) has approximately the same predictions, but ignoring the end points at which two models have  same 
predictions, the ACD model predictions are $1/5$, $5$ and approximately $9$ times of the SM predictions for $P_{LL}$,  $P_{TT}$ and $P_{NN}$, respectively.

\section{Conclusion }
We have calculated some observables such as the branching ratio and double-lepton polarization asymmetries associated with the $B \rightarrow K_0^*(1430) l^+ l^-$
 channel in the framework of the  ACD model with a single universal extra dimension. We discussed the sensitivities of these observables to the compactification parameter, $1/R$. 
We compared the results obtained from the ACD model with the predictions of the standard model. The predictions of the two models approach each other at around  $1000 ~GeV$ for
the value of the compactification parameter, $1/R$. However the results for the two models differ significantly at lower values of the compactification parameter. This difference 
grows specially when the errors of the form factors calculated from the QCD sum rules in Table 1 are taken into account. The maximum deviation from the predictions of SM  for the considered 
quantities, obtained using the central values of form factors, belongs to the case of ACD model predictions, where the errors of the form factors are added to the central values of the form factors.
This deviation also increases as $1/R$ decreases, such that at low $1/R$, the discrepancy between predictions of the SM and ACD models reaches to approximately  one order of magnitude for some 
observables. These results beside the other evidences for deviation of the ACD model predictions from the SM obtained via investigating many observables related to the $B$ and $\Lambda_b$ channels, which due to the heavy bottom quark have
 large range of
$q^2$,  in \cite{buras1,buras2,colangelo0604029,azizi1,colangelo,aliev,aslambey,aslam,colangelo1,Haisch,Mohanta,deste,deste1}, can be 
considered as a signal for the existence of extra dimensions in the nature which should we search for in
  the experiments.

The order of magnitude for the branching ratio shows a possibility to study $B \rightarrow K_0^*(1430) l^+ l^-$ channel at the LHC. Any measurements on the branching ratio as well as the 
double-lepton polarization asymmetries and determination of their signs and their comparison with the obtained results in this paper can give valuable information about
 the nature of the scalar meson $K_0^*(1430)$ as well as the possible extra dimensions.

\begin{figure}
\includegraphics[scale=0.5]{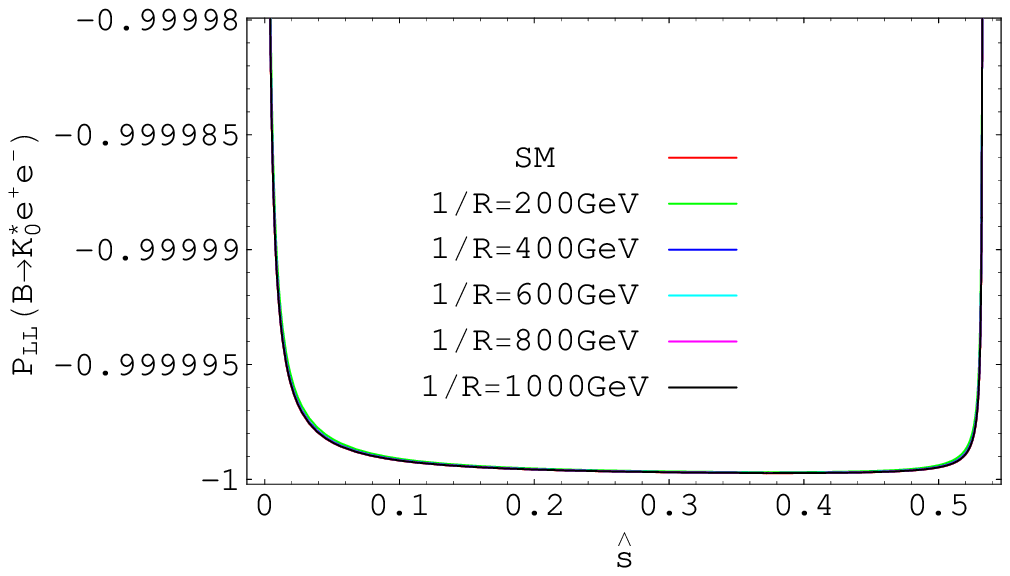}
\includegraphics[scale=0.5]{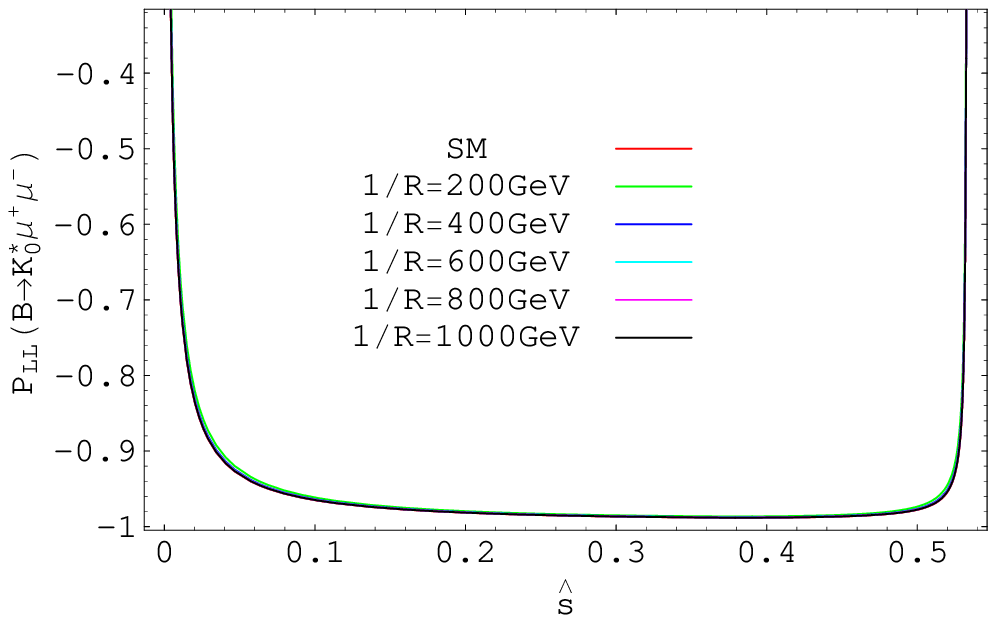}
\includegraphics[scale=0.5]{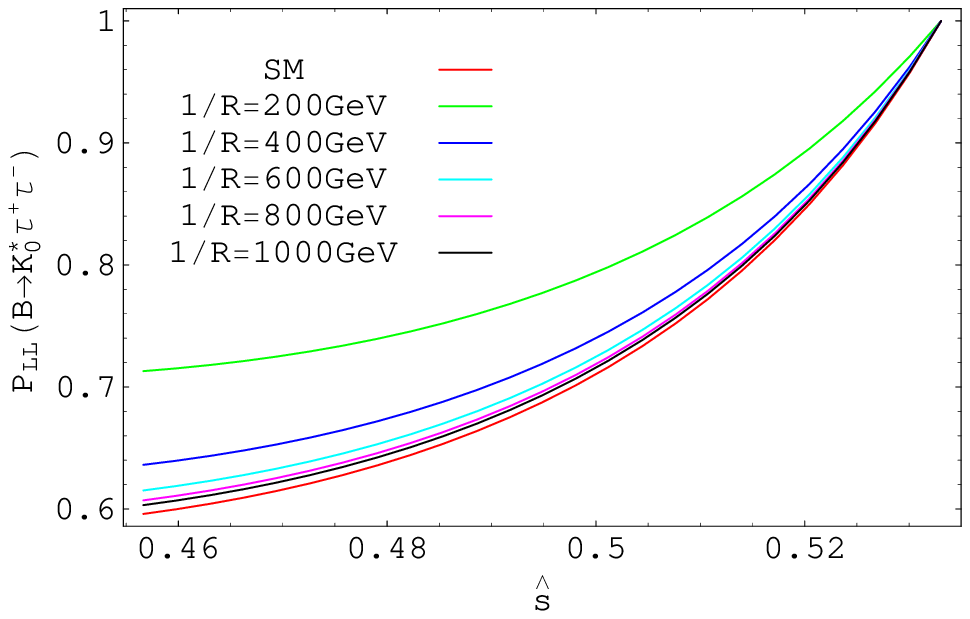}
\caption{ The dependence of the $P_{LL}$ polarization  of the $B
\rar K_{0}^{*} l^+ l^- $ on   $\hat{s} $ at different fixed values of  $1/R$ and the SM for different leptons.\label{figbelma}}
\end{figure}
\begin{figure}
\includegraphics[scale=0.5]{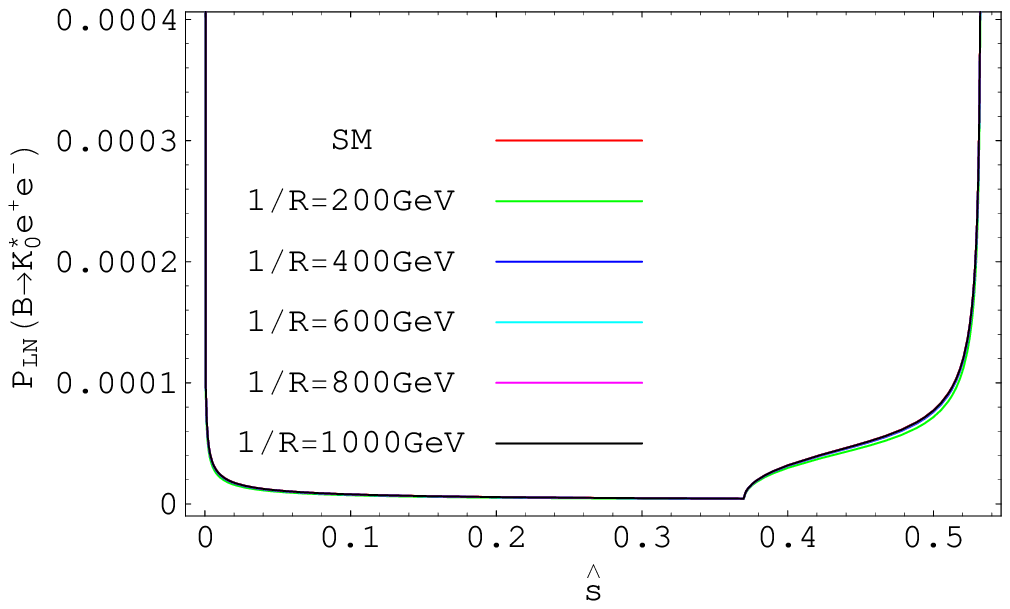}
\includegraphics[scale=0.5]{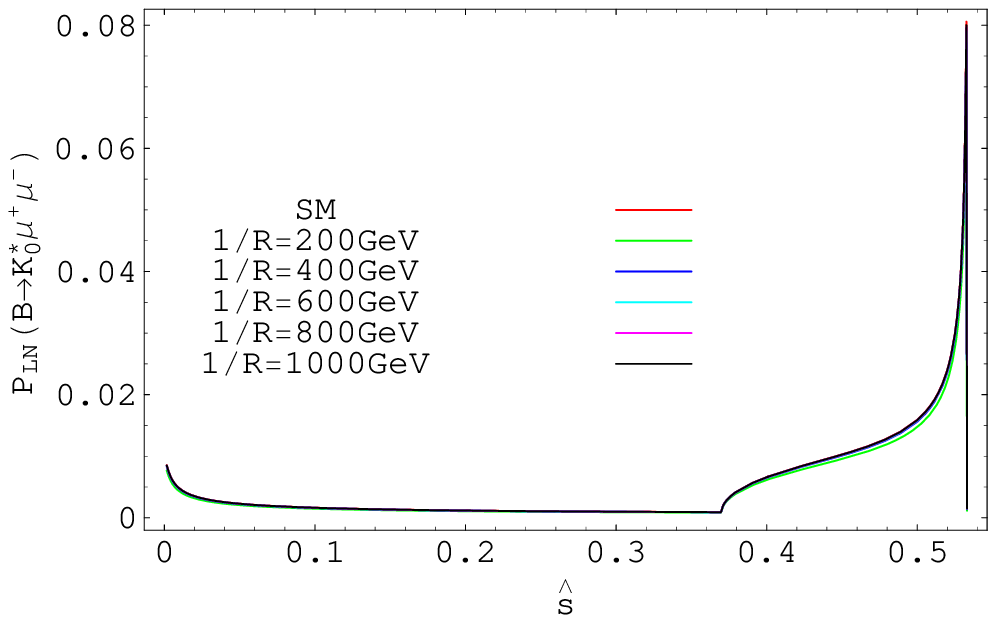}
\includegraphics[scale=0.5]{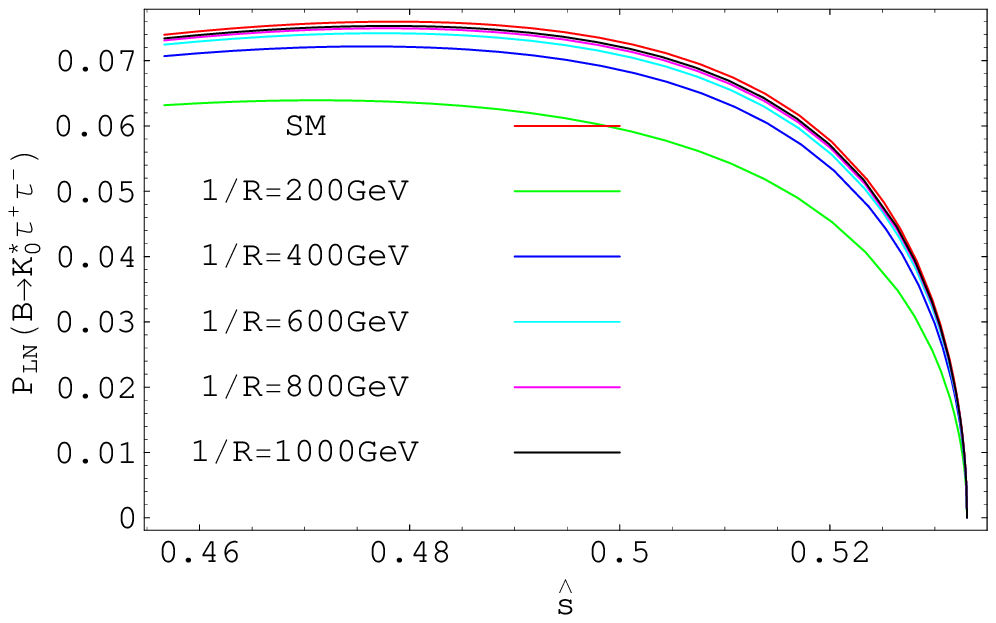}
\caption{ The same as Fig. \ref{figbelma}, but for  $P_{LN}$ polarization. \label{fig4}}
\end{figure}
\begin{figure}
\includegraphics[scale=0.5]{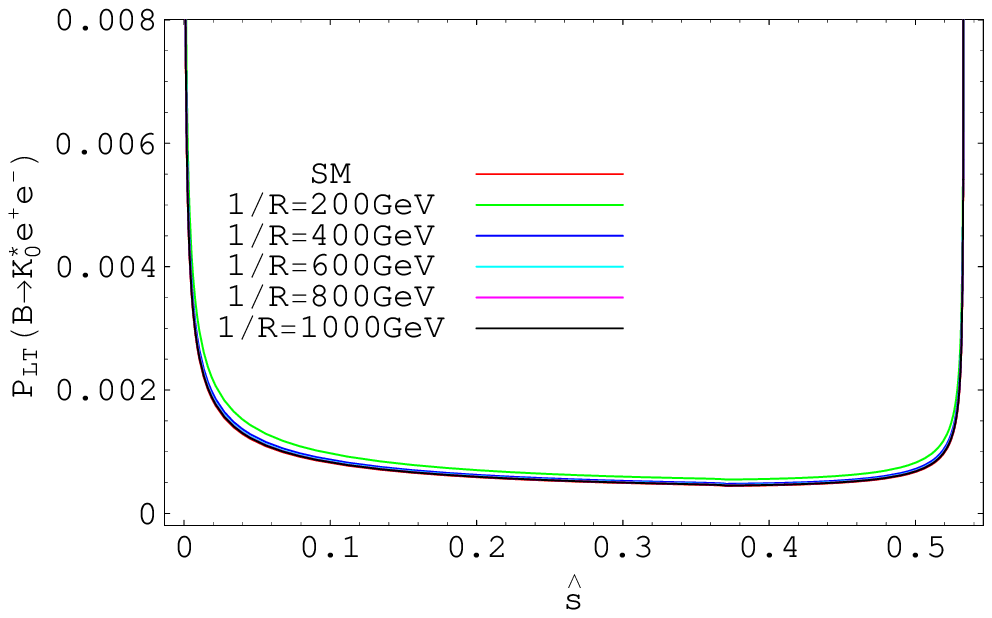}
\includegraphics[scale=0.5]{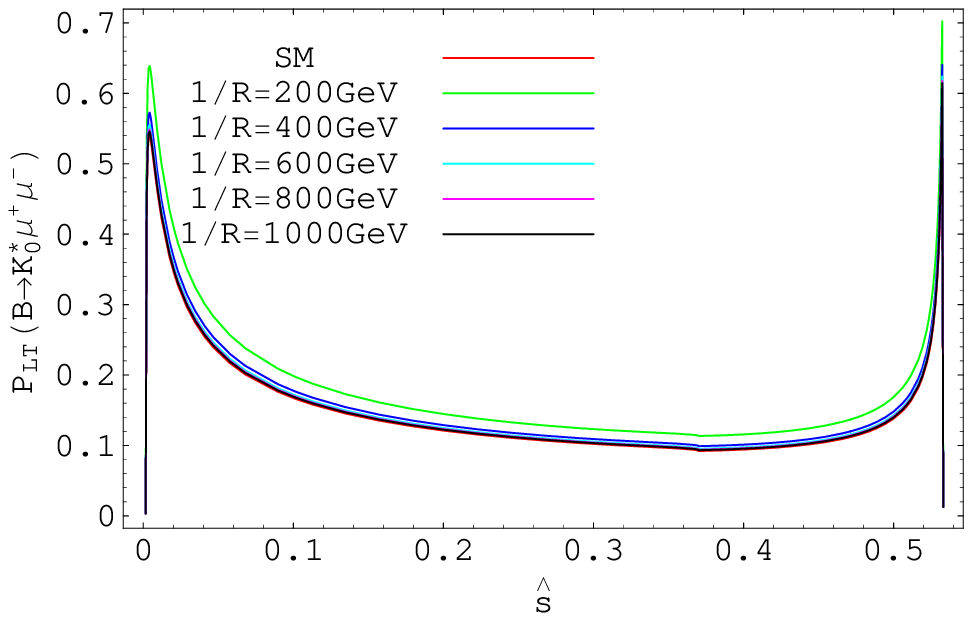}
\includegraphics[scale=0.5]{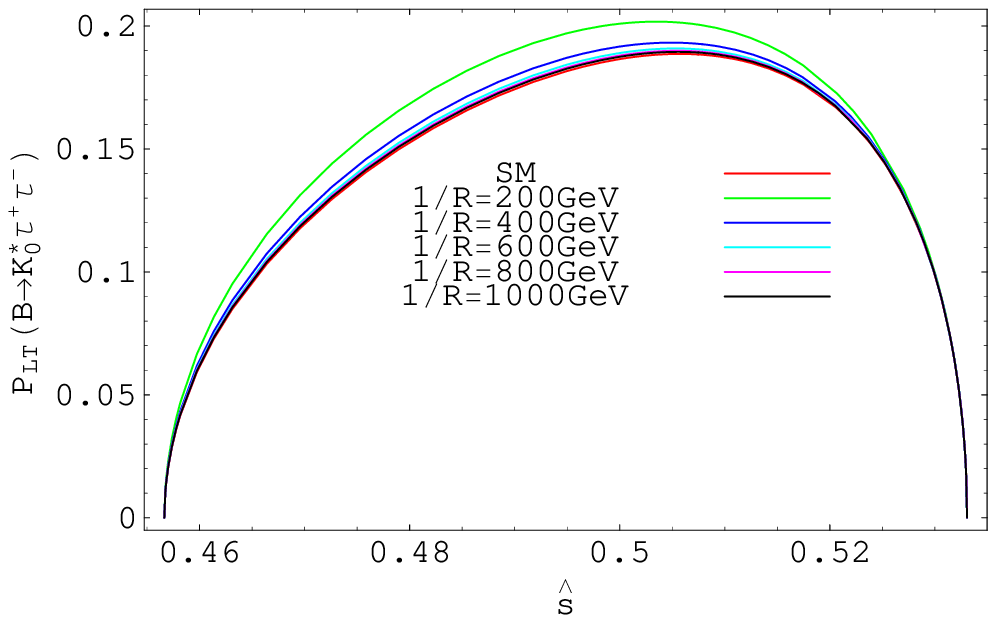}
\caption{The same as Fig. \ref{figbelma}, but for  $P_{LT}$ polarization.  \label{fig6}}
\end{figure}
\begin{figure}
\includegraphics[scale=0.5]{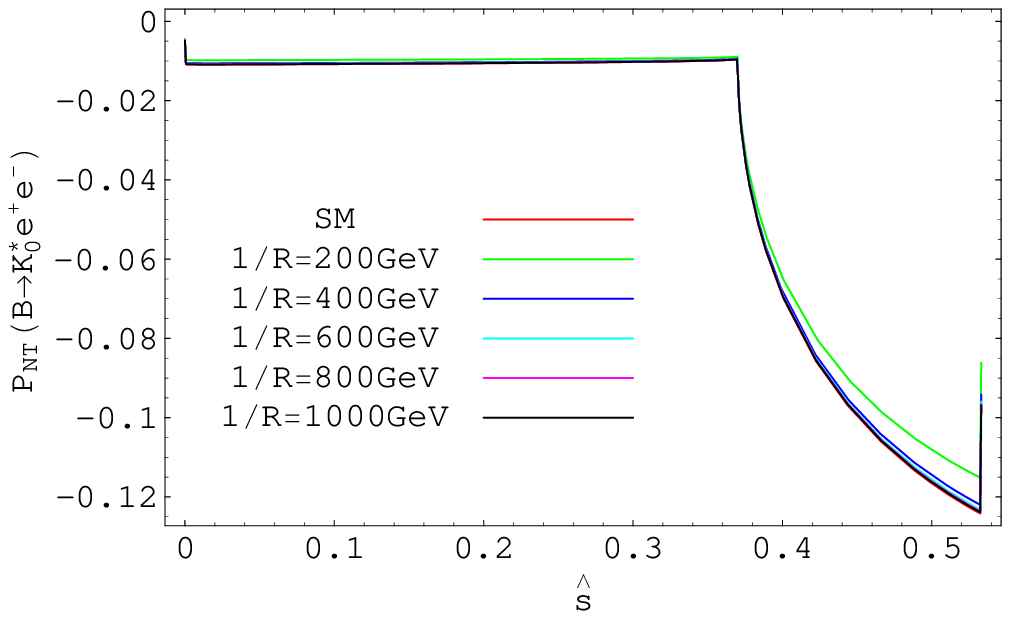}
\includegraphics[scale=0.5]{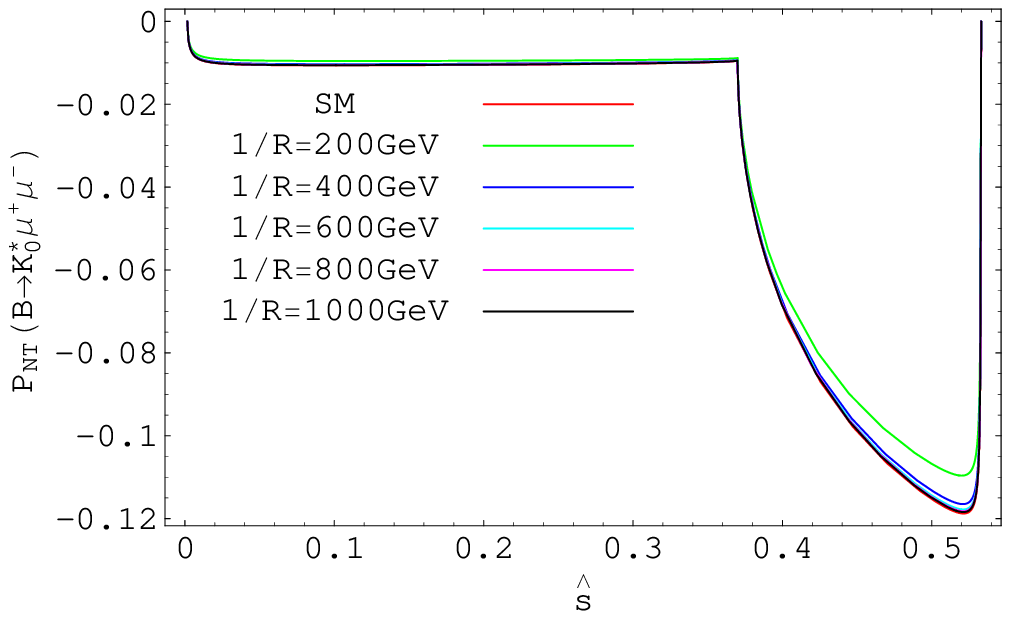}
\includegraphics[scale=0.5]{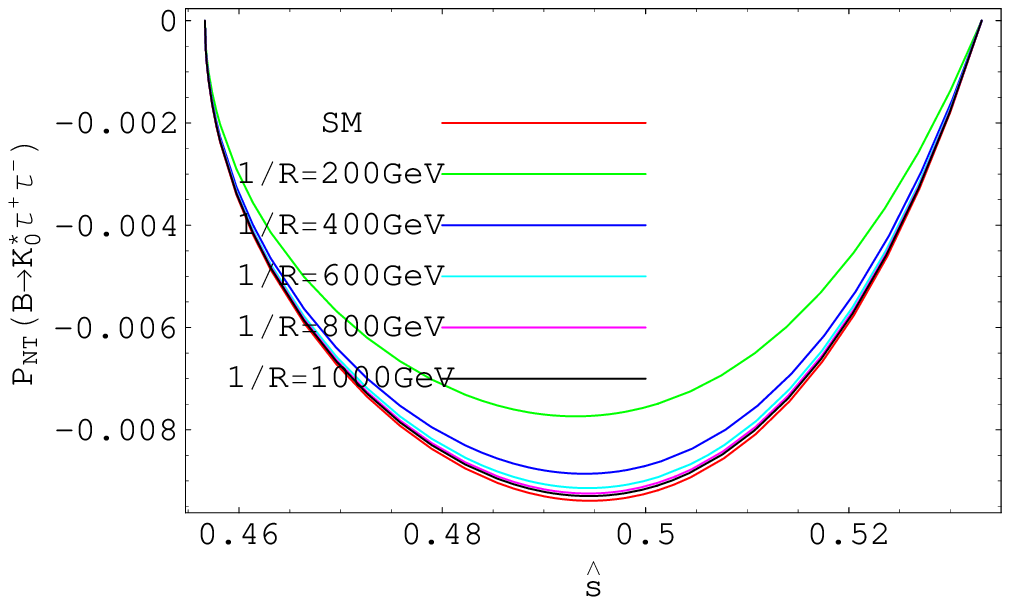}
\caption{ The same as Fig. \ref{figbelma}, but for  $P_{NT}$ polarization.\label{fig8}}
\end{figure}
\begin{figure}
\includegraphics[scale=0.5]{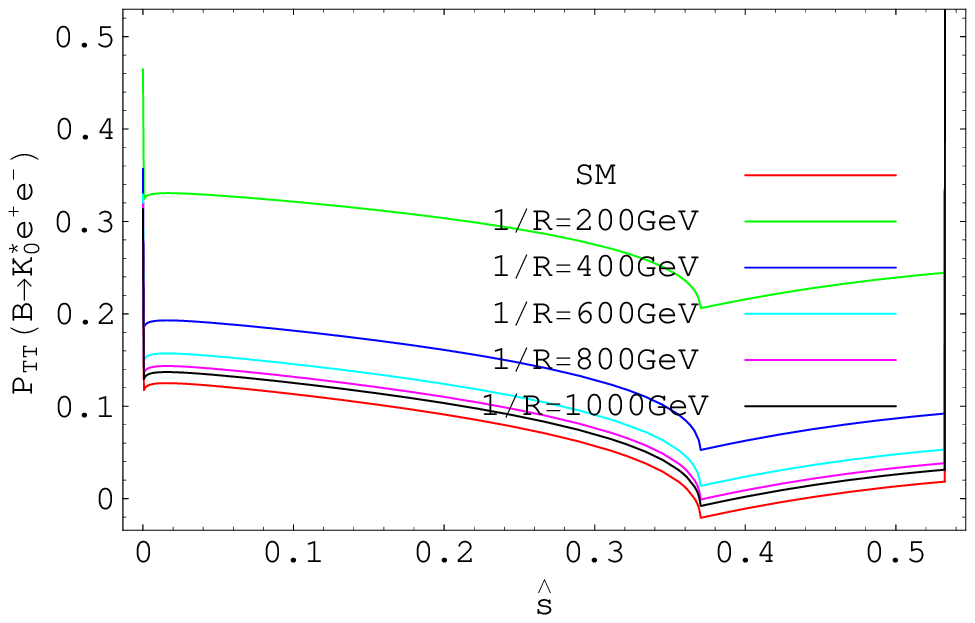}
\includegraphics[scale=0.5]{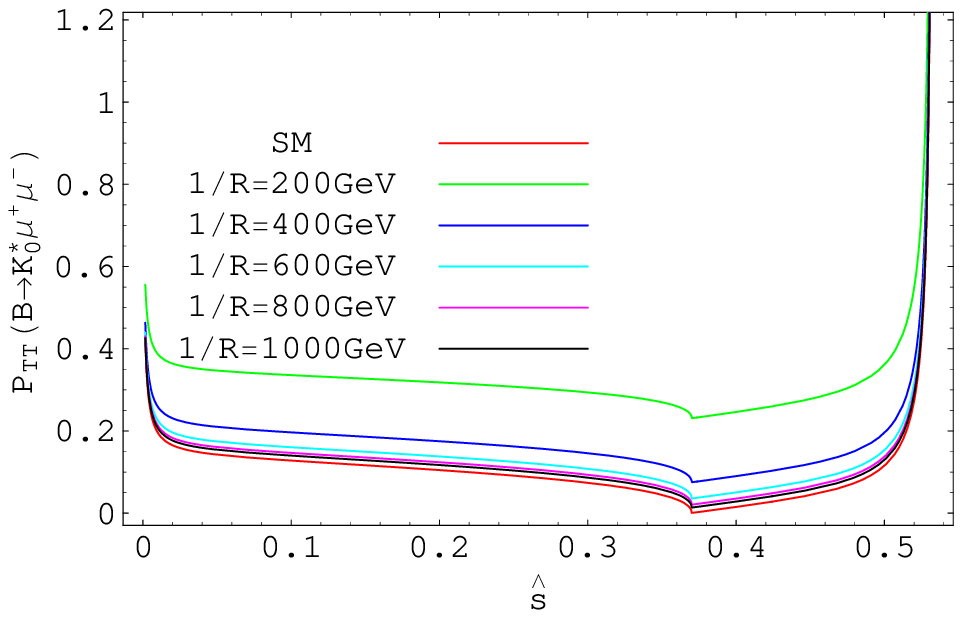}
\includegraphics[scale=0.5]{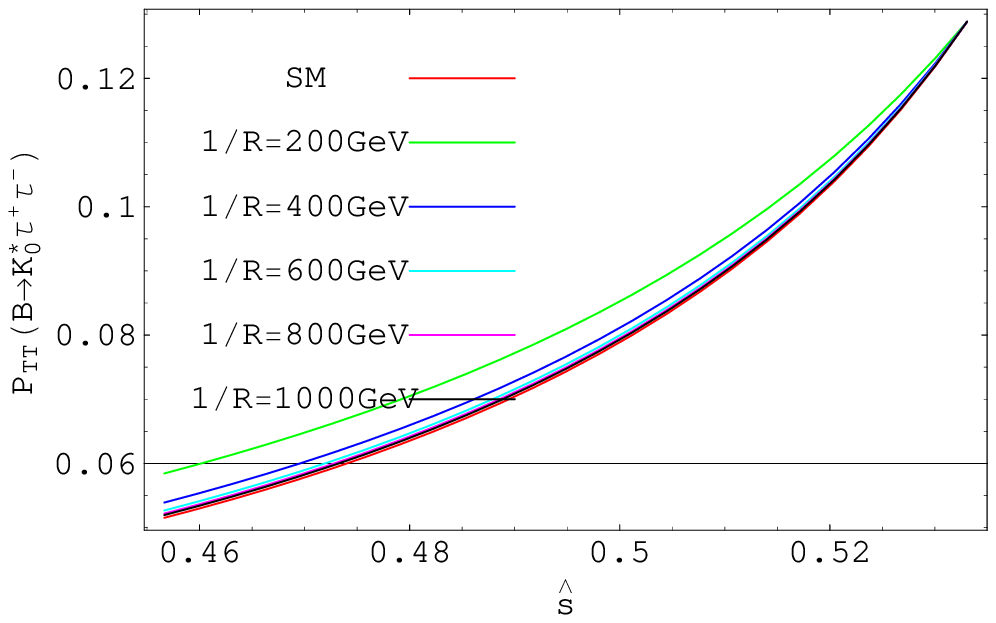}
\caption{The same as Fig. \ref{figbelma}, but for  $P_{TT}$ polarization.  \label{fig10}}
\end{figure}

\begin{figure}
\includegraphics[scale=0.5]{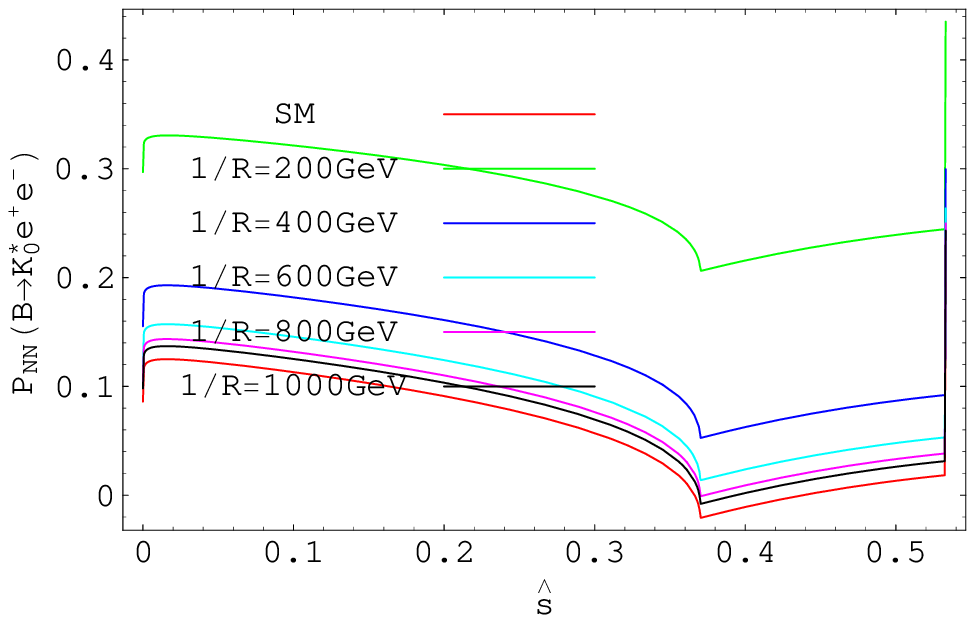}
\includegraphics[scale=0.5]{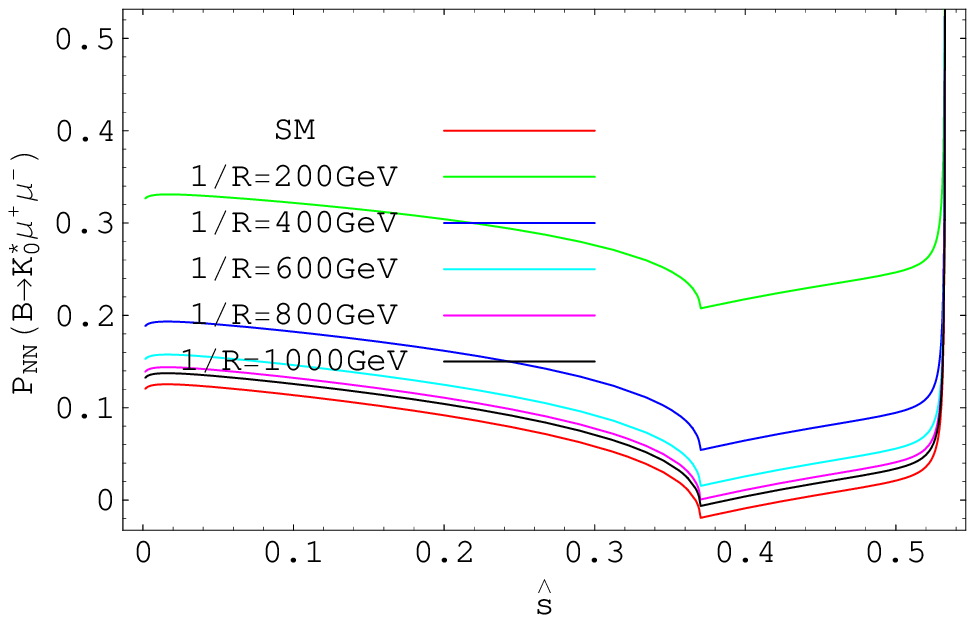}
\includegraphics[scale=0.5]{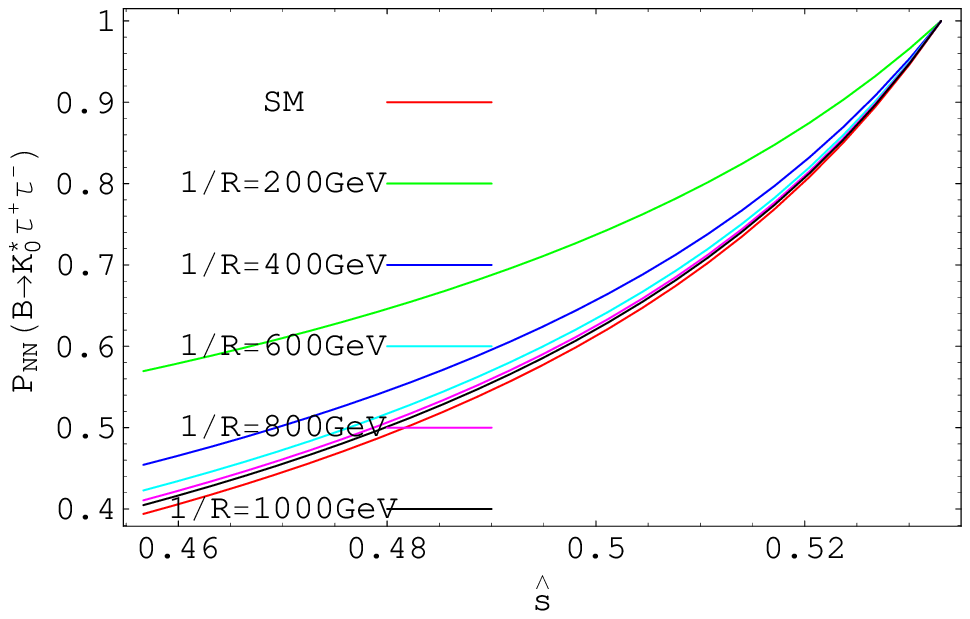}
\caption{The same as Fig. \ref{figbelma}, but for  $P_{NN}$ polarization. \label{fig12}}
\end{figure}
\begin{figure}
\includegraphics[scale=0.75]{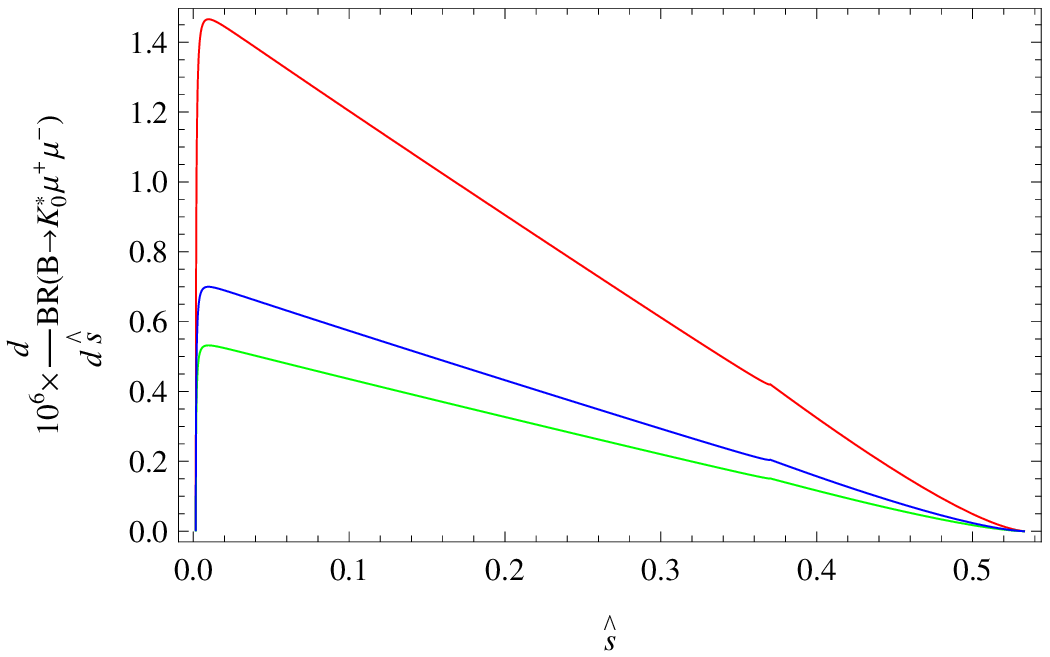}
\includegraphics[scale=0.75]{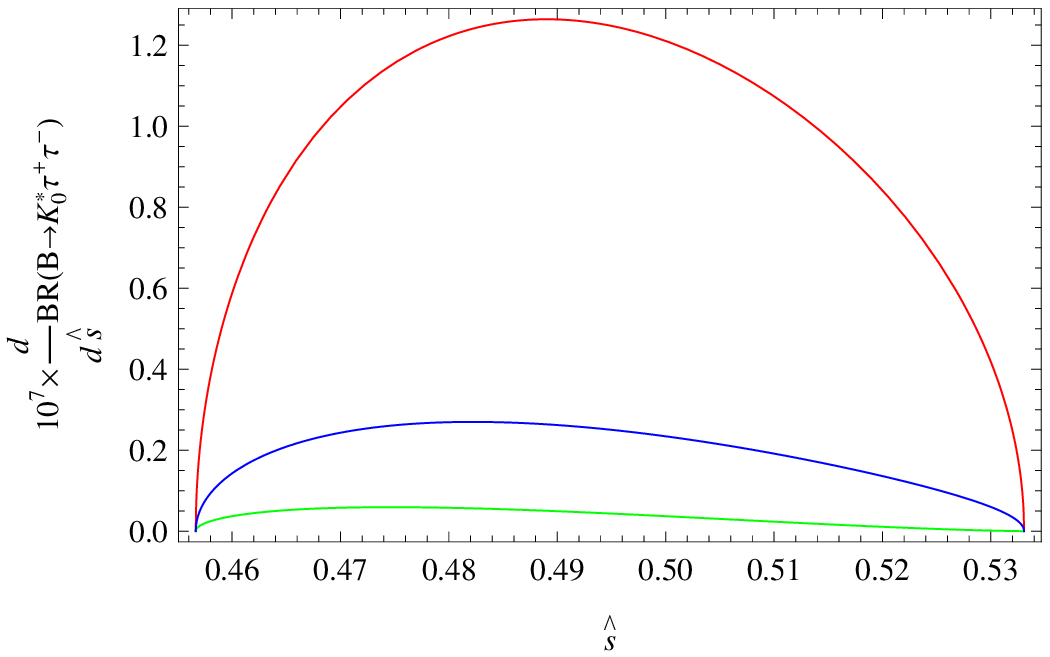}
\caption{The dependence of  differential branching ratio for $\mu$ and $\tau$ on $\hat{s}$.  The red color shows the results at $1/R=200~GeV$ when the errors of form factors are 
added to the central values, the green one shows the predictions at the same value of $1/R$, but when the errors are subtracted from the central values and the blue color shows 
the result of the SM, when the central value of the form factors are considered\label{difbr}}
\end{figure}
\begin{figure}
\includegraphics[scale=0.5]{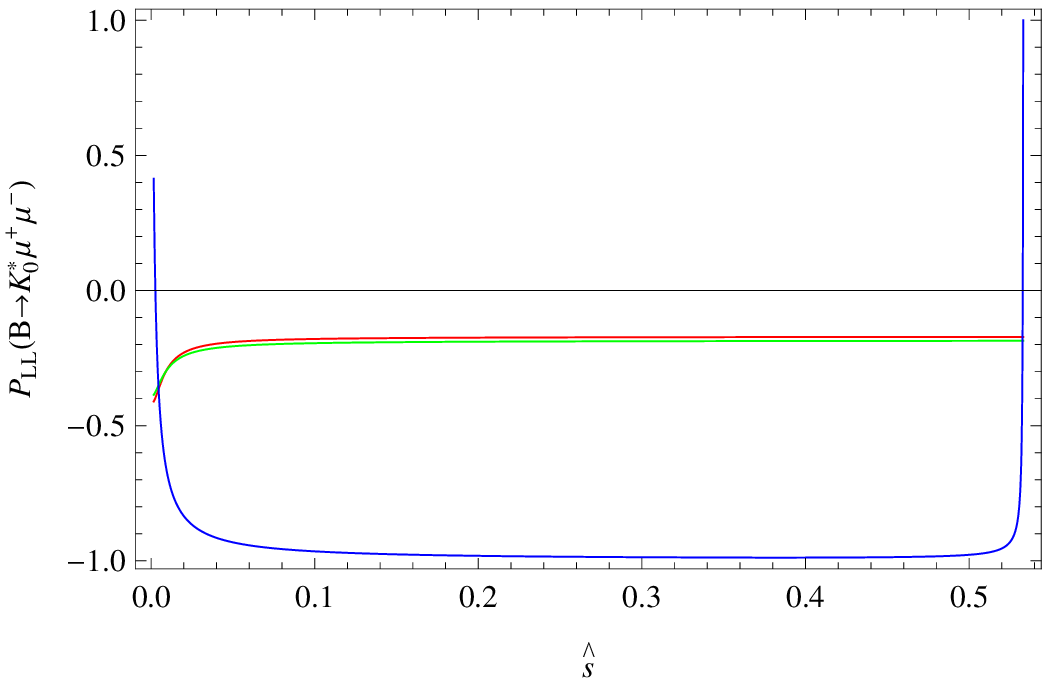}
\includegraphics[scale=0.5]{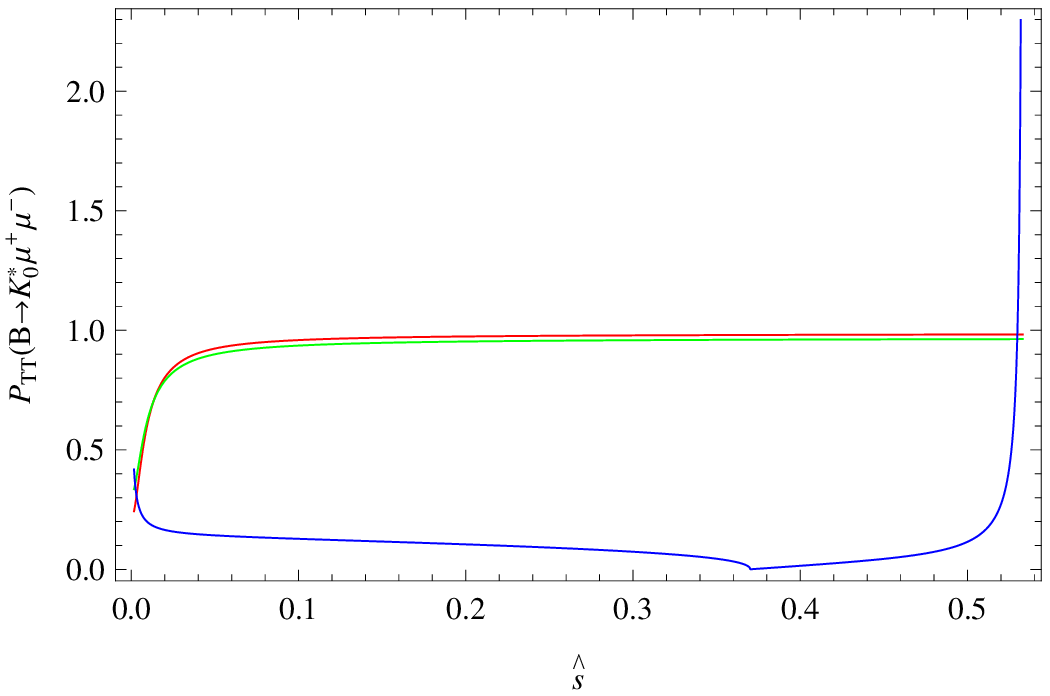}
\includegraphics[scale=0.5]{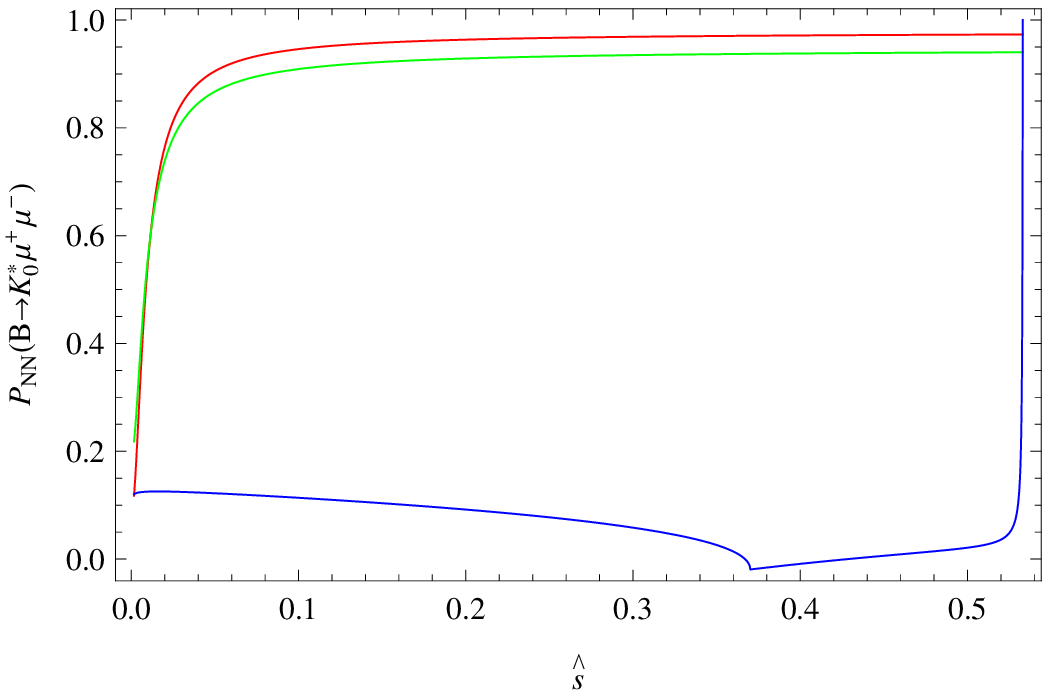}
\caption{The same as Fig. \ref{difbr}, but for $P_{LL}$,   $P_{TT}$  and  $P_{NN}$  polarizations and only for $\mu$. \label{error}}
\end{figure}

%
%
\end{document}